\documentclass[twocolumn,prb,showpacs,amsmath,amssymb,floatfix]{revtex4}
\usepackage{bm}
\usepackage{graphicx}
\usepackage{verbatim}

\usepackage{color}

\begin{document}
\title{ Spin-Orbit Effects in Carbon-Nanotube Double Quantum Dots}
\author{S.~Weiss$^1$,  E.~I.~Rashba$^{2,3,4}$, F.~Kuemmeth$^{2}$, H.~O.~H.~Churchill$^{2}$, K.~Flensberg$^1$}
\affiliation{$^1$ Niels Bohr Institute \& Nano-Science Center, University of Copenhagen, Universitetsparken 5, 2100 Copenhagen, Denmark\\
$^2$ Department of Physics, Harvard University, Cambridge, Massachusetts 02138, USA\\
$^3$ Center for Nanoscale Systems, Harvard University, Cambridge, Massachusetts 02138, USA\\
$^4$ and Department of Physics, Loughborough University, Leicestershire LE11 3TU, UK}
\date{\today}
\begin{abstract}
We study the energy spectrum of symmetric double quantum dots
in narrow-gap carbon nanotubes with one and two
electrostatically confined electrons in the presence of
spin-orbit and Coulomb interactions. Compared to GaAs quantum
dots, the spectrum exhibits a much richer structure because of
the spin-orbit interaction that couples the electron's isospin
to its real spin through two independent coupling constants. In
a single dot, both constants combine to split the spectrum into
two Kramers doublets, while the antisymmetric constant solely
controls the difference in the tunneling rates of the Kramers
doublets between the dots. For the two-electron regime, the
detailed structure of the spin-orbit split energy spectrum is
investigated as a function of detuning between the quantum dots
in a $22$-dimensional Hilbert space within the framework of a
single-longitudinal-mode model. We find a competing effect of
the tunneling and Coulomb interaction. The former favors a
left-right symmetric two-particle ground state, while in the
regime where the Coulomb interaction dominates over tunneling,
a left-right antisymmetric ground  state is found. As a result,
ground states on both sides of the $(11)$-$(02)$ degeneracy
point may possess opposite left-right symmetry, and the
electron dynamics when tuning the system from one side of the
$(11)$-$(02)$ degeneracy point  to the other is controlled by
three selection rules (in spin, isospin, and left-right
symmetry).  We  discuss implications for the spin-dephasing and
Pauli blockade experiments.

\end{abstract}
\maketitle
\section{Introduction}
Coherent control over the charge and/or the spin of an electron
or hole is a key ingredient for quantum computation or
spintronic devices. It is of importance to have coupled two
level systems (qubits) that can be controlled and manipulated
efficiently without loss of the stored information. A promising
candidate and natural two-state system for a robust qubit is
the spin of an electron. Coherent manipulation as well as
preparation and read-out of a single confined spin in few
electron semiconductor quantum dot (QD) systems have been
demonstrated, see Refs. \onlinecite{Hanson} and
\onlinecite{Petta} and references therein.

Spin qubits in carbon nanotubes\cite{Iijima, Kuemmeth2} (CNT)
are believed to be even more robust due to the absence of
hyperfine coupling in $^{12}C$.\cite{Churchill2} This is in
contrast to GaAs QDs where the phase coherence suffers from
hyperfine coupling due to the nuclei of the host crystal.
However, CNTs pose other challenges and complications. First,
few electron QDs are not easily fabricated and, secondly, the
isospin degree of freedom present in the honeycomb carbon
lattice provides another quantum two level system that must be
included in the analysis.

The band structure of electrons in nanotubes can be understood
starting from that of graphene,\cite{Dresselhaus,Castro} which
has a linear dispersion relation similar to massless Dirac-Weyl
fermions. Graphene is a zero gap semiconductor, but when the
graphene sheet is rolled to form a nanotube, the quantization
condition for nanotubes leads to metallic or semiconducting
behavior, depending on chirality. \cite{Ando,Roche} The curved
geometry creates a mass term in the Dirac spectrum and thus a
bandgap even for the nominally metallic tubes.\cite{Kanecurv,
Kleiner, Mintmire, Hamada, Saito1} This bandgap allows for
electrostatic confinement of electrons and creation of few
electron QDs, otherwise not possible due to the Klein
paradox.\cite{Kouwenhoven} Recent experiments have shown that
it is indeed possible to confine electrons in single
\cite{Jarillo,Minot, Kuemmeth,Leturcq} and double quantum dots
(DQD) in a CNT  by means of electrostatic gates in cleanly
grown small bandgap nanotubes.\cite{Churchill,
Kouwenhoven,Churchill2} The present study is motivated by these
experimental results.

In the recently investigated few-electron nanotube QDs, the
four-fold degeneracy due to the spin and isospin degrees of
freedom is split by SO coupling, giving rise to  a coupling of
spin and isospin degrees of freedom. In plane graphene,
spin-orbit (SO) coupling (being a relativistic effect) of its
$\pi$-electrons is of dozens of $\mu$eV only \cite{MacDonald,
Yao,Boettger,Gmitra} and therefore of minor importance. The
curved geometry of nanotubes induces SO coupling on the order
of $10^{-1}$ meV among the single particle levels of the
electrons.\cite{Kuemmeth} While the curvature-induced SO
coupling was envisioned previously for
semiconductors,\cite{Entin,Bulgakov} for nanotubes it is the
dominant mechanism. It was Ando \cite{AndoSO} and others \cite{DeMartino,Chico2004}
who developed the first theories of SO coupling in nanotubes. More recent
theoretical investigations extended this work by including the
$\sigma$- and $\pi$- bands in full as well as the curved bonds
between neighboring atoms.\cite{HH,Izumida, Jeong, Chico}
Lowest order perturbation theory shows that the SO coupling is
inversely proportional to the radius of curvature and
originates from the intra-atomic SO coupling in a carbon atom.
Even though this is a weak coupling compared to heavier atoms,
the combined effect of curvature and intra-atomic SO coupling
splits a four-fold degenerate level into two Kramers doublets
by a fraction of a meV. The demonstration of electrostatically
confined particles in CNT-QDs in the presence of SO coupling
\cite{Kuemmeth,Churchill} has motivated recent theoretical
investigations. The single electron QD setup and in particular
the influence of the electron-phonon coupling on the
decoherence are subjects of a work by Bulaev et. al.
\cite{Bulaev} The two-particle problem has been studied
numerically in Refs.~\onlinecite{Wunsch} and
\onlinecite{Rontani} for a hard wall and a harmonic potential,
respectively. Furthermore, hyperfine interactions and their
consequences for Pauli blockade have been discussed.
\cite{Fischer,Burkard}

Here, we present a theory of the energy spectrum for a CNT-DQD
in the presence of SO coupling in the envelope function
formalism within an exact diagonalization scheme. The system is
described by three quantum numbers [spin $s$, isospin $\tau$,
and left and right dots ($L/R$)], each taking two  values, and
by the discrete and continuum spectra of the longitudinal
motion. Depending on the specific parameter values of the DQD
that vary in a wide range, two electrons confined in a DQD find
themselves in rather different regimes. In this article, we
concentrate on coated narrow-gap nanotubes of the kind
investigated in Refs.~\onlinecite{Churchill2} and
\onlinecite{Churchill} that seem well suited for spintronic
applications. In such DQDs, a small electron mass $m^*$
increases the separation between the levels of longitudinal
quantization, while a high dielectric constant $\kappa\sim10$
suppresses the Coulomb repulsion, which would otherwise result
in level mixing. Restricting ourselves with the lowest
longitudinal mode, we concentrate instead on the detailed
structure of the energy spectrum emerging from the spin-isospin
coupling, and the dependence of the interdot tunneling and
Coulomb energies on the symmetry of wave functions and SO
coupling. However, we stop short of discussing the influence of
phonon and hyperfine couplings as well as the scattering
between different isospin states.\cite{Burkard} It
should be mentioned that similar physics occurs in silicon
double quantum dots where, however, the valley degeneracy is
usually broken and spin-orbit interaction does not play an
important role.\cite{Culcer}

To set up our model calculation, we use the well-established
model for $p_z$-orbitals of graphene to describe electrons in a
nanotube. The DQD confinement is modeled by a double
square-well potential along the axial direction of the
nanotube. We take into account SO coupling effects on the
single particle levels and discuss their influence on the
energy spectrum in the presence of an axial magnetic field $B$.
In the framework of a single-mode approximation, and using the
eight-function basis $(8=2\times2\times2)$ of single-electron
states, we present a symmetry classification of the eigenstates
of a two-electron symmetric DQD in its 22-dimensional Hilbert
space. We find its energy spectrum numerically for the
comparable values of the tunneling integral, the Coulomb
interaction and the SO splitting. Our main results include the
detuning dependence of the spectrum and the effect of a
magnetic field lifting all spectrum degeneracies. Finally, we
discuss challenges and opportunities for experimental studies.

The structure of the article is as follows. After we have
summarized the physical properties of a single electron in a
nanotube in Sec. \ref{hamiltonian}, we turn in Sec.
\ref{DQDsol} to the model of an electrostatically generated DQD
and solve the eigenvalue problem for a  DQD with a square-well
potential. We clarify the effect of two SO coupling constants,
$\Delta_0$ and $\Delta_1$, on the energy splitting between the
Kramers doublets and on the tunneling integral. Sec.
\ref{twopart} presents the symmetry classification of the
two-electron wave functions as well as a description of the
techniques used for calculating Coulomb integrals on SO
modified wave functions, and the effect of SO coupling on
Coulomb integrals as well as their $B$-dependence. In
Sec.~\ref{detun} we present our main results on the energy
spectrum of a two-electron DQD as a function of detuning and
its transformation when a magnetic field is applied. Summary
and discussion are presented in Sec.~\ref{conclusion}.

\section{Model}
\subsection{Single electron in a nanotube}
\label{hamiltonian}
\begin{figure}[t!]
\begin{center}	
\includegraphics[width=0.5\textwidth]{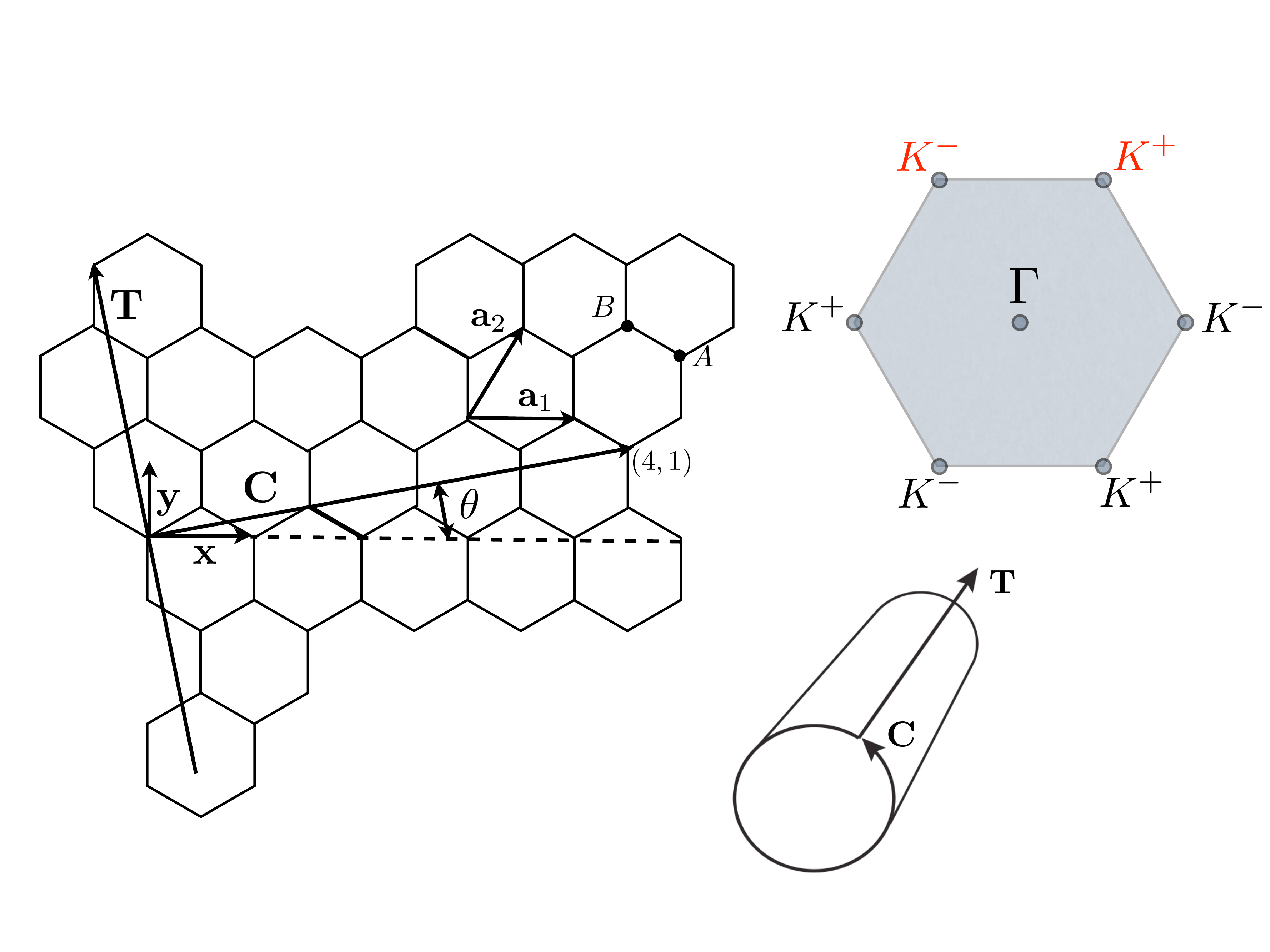}
\caption{ \label{honeycomb} Sketch of the planar graphene sheet showing the honeycomb lattice structure. Nearest neighbors belong to two different sublattices $A$ and $B$. A nanotube with chirality $(4,1)$ is formed when the sheet is rolled up along the direction of the chiral vector ${\bf C}$. The chiral angle $\theta$ gives the misalignment between the chiral vector and the primitive lattice vector ${\bf a}_1$. The direction perpendicular to the chiral vector defines the tube axis and is denoted ${\bf T}$. Within the tight-binding approximation, we assume the $z$-direction to be perpendicular to ${\bf T}$ and ${\bf C}$, i.e., the $p_z$ orbitals stick out of the plane of the figure. The Brillouin zone
of the honeycomb lattice is shown in the upper right corner.
The Dirac points are denoted by ${\bf K}^{\tau}$, and we choose
in our calculations two inequivalent points $K^{\tau}=2\pi/a(\tau/3,1/\sqrt{3})$, which are marked in red. A graphene sheet folded into a nanotube is shown in the lower right corner.
}
\end{center}
\end{figure}

We consider a single-wall CNT whose electronic properties are
described within a tight-binding model for the $p_z$-orbitals
of neighboring carbon atoms \cite{Dresselhaus}. As usually, we
solve for the band structure of a plane graphene sheet first
and then impose periodic boundary conditions for the electronic
motion along the circumferential direction defined by a chiral
vector ${\bf C}$.  It is defined as ${\bf C}=n_1{\bf
a}_1+n_2{\bf a}_2$ in terms of the primitive lattice vectors
${\bf a}_1=a(1,0)$ and ${\bf a}_2=a(1/2, \sqrt{3}/2)$, where
$(n_1,n_2)$ are integers. Coordinates along the circumferential
and translational directions, $\bf C$ and $\bf T$ in
Fig.~\ref{honeycomb}, are $(c,t)$. Due to the honeycomb lattice
structure of graphene, nearest-neighbor atoms belong always to
different sub-lattices $A$ and $B$; the lattice constant is
$a=0.246$ nm. The chiral angle, which is the angle between
${\bf C}$ and ${\bf a}_1$, is
$\theta=\arctan[\sqrt{3}n_2/(2n_1+n_2)]$. In graphene, two
spin-degenerate $\pi$-bands (the conduction and valence bands)
cross at six vertices of the Brillouin zone. Two pairs of
translationally nonequivalent vertices, ${\bf K}^\tau$, form
two Dirac points; therefore graphene is a semi-metal. Here we
choose the Dirac points  as  ${\bf
K}^{\tau}\equiv2\pi/a(\tau/3,1/\sqrt{3})$, with $\tau=\pm 1$.
The effective low-energy Hamiltonian is obtained by expanding
in the electron momentum near the Dirac points ${\bf K}^\tau$,
\cite{Roche,Ando,KaneMele}
\begin{equation}
\mathcal{H}=\hbar v(\tau_3k_c\sigma_1+k_t\sigma_2). \label{graphene}
\end{equation}
Here, $\tau_3$ is the diagonal Pauli matrix with eigenvalues $\tau=\pm 1$
in the ${\bf K}^{\tau}$-isospin subspace.
The Pauli matrices $\sigma_j~ (j=1,2)$ act in sublattice space and account for the two carbon atoms in the primitive unit cell of the honeycomb lattice (up to a unitary transformation that depends on $\theta$, see Ref.~\onlinecite{Bulaev}). The quasimomentum components along the $\bf C$ and $\bf T$ axes
are $k_c$ and $k_t$, see Fig.~\ref{honeycomb}. The eigenvalues of the Hamiltonian $\cal H$
 are readily obtained
\begin{equation}
E_{k_c,k_t}=\pm\hbar v\sqrt{k_c^2+k_t^2}, \label{engraph}
\end{equation}
where $\pm$-solutions in Eq.~(\ref{engraph}) correspond to the conduction and valence band, respectively, that  are degenerate in the isospin quantum number
$\tau=\pm1$, and $v\approx 8\times10^5$ m/s is the Fermi velocity
in graphene. Plane-wave type eigenfunctions for the Hamiltonian $\cal H$ are
\begin{equation}
\Psi^{\tau}_{k_c,k_t}(c,t)=\frac{e^{i{\bf K^\tau}\cdot{\bf r}}}{\sqrt{4\pi}}\exp\left\{i(k_c c+k_tt)\right\}\left(\begin{array}{c}z^{\tau}_{k_c,k_t,c/v}\\ 1\end{array}\right), \label{elnt}
\end{equation}
where the coefficients $z^{\tau}_{k_c,k_t,c/v}$ for the conduction or valence band (denoted by the subscripts $c$ and $v$) are
\begin{equation}
z^\tau_{k_c,k_t,c}=\frac{\tau k_c- ik_t}{\sqrt{k_c^2+k_t^2}}, \hspace{1cm} z^\tau_{k_c,k_t,v}=-\frac{\tau k_c- ik_t}{\sqrt{k_c^2+k_t^2}}. \label{zfunc}
\end{equation}
The position vector for the electron is ${\bf r}={\bf
r}(x,y)=(c\cos\theta-t\sin\theta,c\sin\theta+t\cos\theta)$ and
the radius of the tube is $R=|{\bf
C}|/2\pi=a\sqrt{n_1^2+n_2^2+n_1n_2}/2\pi$. For
$(n_1,n_2)=(n,0)/(n,n)$ a CNT is called zigzag/armchair like,
other nanotubes are called chiral.\cite{Roche, Ando}
Obviously, for zigzag tubes $\theta=0$ and $(x,y)\equiv (c,t)$.
In what follows, we restrict ourselves with conduction band
electrons and designate amplitudes as $z^\tau_{k_c,k_t}$.

Wavefunctions of nanotubes are periodic in the circumferential
direction, i.e., $\Psi({\bf r})=\Psi({\bf r+C})$. Consequently,
the wavenumber $k_c$ of electrons/holes in this direction is
quantized by the condition ${\bf (k+K^\tau)\cdot C}=2\pi m$
with $m$ being integer numbers and ${\bf k}=(k_c,k_t)$.
Depending on the direction and magnitude of $\bf C$, there are
two types of solutions obeying the periodicity condition around
the circumference. If we use
\begin{equation}
 e^{i{\bf K}^\tau\cdot{\bf C}}=\exp\left\{(2\pi i\tau/3)(n_1-n_2)\right\}
\label{period}
\end{equation}
as well as $(n_1-n_2)=3M+\nu$ with $\nu=0,\pm 1$ and $M$
integer, we obtain the quantization of the wavenumber around
the circumference as $k_c\to k_m\equiv (m-\nu\tau/3)/R$. Note
that $\nu=0$ is always fulfilled for armchair tubes, but for
zigzag tubes only if $n_1=3M$.  For $\nu=\pm 1$, the envelope
wavefunctions accumulate phase factors.

In graphene, a classification of quantum states by ${\bf
K}^\tau$ is protected by the conservation of the crystal
momentum $\bf k$. In nanotubes, however, it is at first sight
not so clear that isospin is a good quantum number, since for
some metallic tubes the folding of the graphene bandstructure
onto the first Brillouin zone of the translational nanotube
unit cell, results in Fermi points at $k_t=0$ for both isospin
values (sometimes classified as ``zigzag-like"
nanotubes\cite{Lunde} or in Ref.~\onlinecite{Dresselhaus}
chapter 4 as metal-1). Nevertheless, for these nanotubes the
isospin quantum number is protected by a screw axis of the
order $N_S(n_1,n_2)$ defined by a Diophantine equation, see
Ref.~\onlinecite{White}. For the $k_t=0$ point, screw rotations
are equivalent to spatial rotations, hence, it follows from
Eq.~(\ref{period}) that such rotations produce phase factors
$\exp[(2\pi i\tau/3N_S)(n_1-n_2)]$ having complex conjugate
values for $\tau=\pm1$. Therefore, $\tau=\pm$ states belong to
complex conjugate representations. For the other class of
metallic tubes (``armchair-like" or metal-2) where the Fermi
points are different and at $\pm 2\pi/3T$, isospin is protected
by momentum conservation. This should clarify the meaning of
the isospin quantum number $\tau$ for specific types of
nanotubes.

If we insert the allowed quantized $k_m$ values into the
dispersion relation of Eq.~(\ref{engraph}), we see that there
is a gap between the conduction and valence bands given by
$2E_g$ with  $E_g=\hbar v |m-\nu\tau/3|/R$. For $m=0$ and
$\nu=\pm 1$, this gap between the conduction and valence bands
is about $2E_g\approx360$ meV for $R=1$~nm.  Such nanotubes are
thus semiconducting, whereas the $\nu=0$ tubes are nominally
metallic. The curvature, however, opens a small gap,
\cite{Kleiner, Kanecurv,Mintmire, Hamada, Saito1} likely
causing the measured gaps of order $10-50$ meV in
Refs.~\onlinecite{Kuemmeth} and \onlinecite{Churchill}. The
curvature effects appear in the Dirac Hamiltonian as a mass
term, and  Eq.~\eqref{graphene} reduces to a one-dimensional
effective Hamiltonian with $k_c$ modified as (for the lowest
energy mode $m=0$)
\begin{equation}
k_c\rightarrow \tau  k_g, \,k_g=-{{\nu}\over{3R}}+ k_\mathrm{curv}\,,
\label{kg}
\end{equation}
where the last term scales with tube radius as $k_\mathrm{curv}\propto 1/R^2$ and the induced gap as $E_{\mathrm{curv}}=\hbar vk_\mathrm{curv} \propto 1/R^2$.\cite{Kanecurv,Kleiner} While specific expressions for the gap are model dependent, the order of magnitude estimate for $R$ of a few nanometers is\cite{Kanecurv,Kleiner, Izumida}
\begin{equation}
E_{\rm curv}(\theta)\sim(\hbar va/R^2)\cos3\theta\sim 10 \mbox{ meV}.
\end{equation}

We also include a magnetic field $B$ which points in the tube axis direction ${\bf T}$ and induces an Aharonov-Bohm flux $\Phi_{AB}=B\pi R^2$ through the cross section of the tube. This further modifies the circumferential momentum as  $\tau k_g\rightarrow \tau k_g+\Phi_{AB}/(R\Phi_0)$, with $\Phi_0=hc/e$ being the flux quantum. Therefore, the nonrelativistic circumferential momentum $k_c^{\rm nr}$ equals
\begin{equation}
k_c^{\rm nr}=\tau k_g+{{\Phi_{AB}}\over{R\Phi_0}}.
\label{eqkc}
\end{equation}

Besides the orbital effect, the magnetic field also leads to a Zeeman term given by $S_tg\mu_B B$, where $S_t=\pm1/2$ is the spin projection along the CNT axis, $\mu_B$ is the Bohr magneton, and $g\simeq 2$ is the bare electronic $g$-factor. This yields an energy difference between the different spin species of the electron.  In this paper, we only consider tubes with finite gaps allowing for electrostatic confinement of electrons, and pay special attention to the tubes with the curvature-induced gaps (narrow-gap nanotubes, $\nu=0$). We therefore write the (electron/hole) dispersion relation as
\begin{equation}\label{Ekt}
E_{k_c^{\rm nr},k_t}=\pm\hbar v\sqrt{(k_c^{\rm nr})^2+k_t^2}+S_tg\mu_B B.
\end{equation}

Now we introduce the SO coupling which was shown to be an important effect in the recent experiments on few electron QDs. \cite{Kuemmeth,Churchill} In general, the coupling of the electron spin to its orbital motion is a relativistic effect for electrons moving in external electric fields. Asymmetric confinements in semiconductor QDs (extrinsic SO coupling, see Ref. \onlinecite{vanderWiel} and references therein) can also provide such a coupling which, however, is one order of magnitude smaller than the SO coupling constants reported in CNTs. The pioneering theoretical work on the curvature induced SO derives the low-energy effective Hamiltonian from a first-order perturbation theory in the atomic SO coupling as well as in the curvature.
\cite{AndoSO} Due to curvature, there are nonzero overlaps between the $p_z$ and $p_{x,y}$ orbitals of neighboring atoms. Combined with the atomic SO coupling which produces transition matrix elements between different quantum states on the same atom, a spin dependent coupling between the adjacent $A$ and $B$ atoms arises. \cite{AndoSO,HH} More recent work \cite{Chico,Izumida,Jeong} has extended this approach and added to the low-energy effective Hamiltonian of $\pi$-electrons in CNTs a term that is diagonal in sublattice $(A,B)$ space.
The generalized SO Hamiltonian near the Dirac points is
\begin{equation}
H_{SO}=\Delta_1\tau_3\sigma_1s_3+\Delta_0s_3\tau_3, \label{HamSO}
\end{equation}
with $s_3$ being a diagonal Pauli matrix acting in spin space. According to Eq.~(\ref{HamSO}), the electron spin is still a good quantum number for the single particle problem, and in what follows the eigenvalues of $s_3$ are denoted as $s=\pm1$. The two coupling constants $\Delta_0$ and $\Delta_1$ depend on the type of the tube. Both are inversely proportional to the radius $R$, and $\Delta_0$ depends on the chiral angle as $\Delta_0\propto\cos(3\theta)$, while $\Delta_1$ does not depend on $\theta$.
 The term in Eq.~(\ref{HamSO}) proportional to $\Delta_{0}$ is diagonal in sublattice space and is responsible for the difference in the electron and hole spin-orbit gaps observed experimentally;\cite{Kuemmeth} see Eq.~\eqref{deltas} below. The combination of the SO interaction with the effective nanotube Hamiltonian described above gives the final expression for the
dispersion relation in the conduction and valence bands
\begin{align}
E_{k_c,k_t}^{\tau,s}=\pm\hbar v\sqrt{k_c^2+k_t^2}
+\left(\Delta_0 \tau+\frac{1}{2}g\mu_B B\right) s,\label{spenergy}
\end{align}
with
\begin{equation}
k_c=\tau k_g+{{\Phi_{AB}}\over{R\Phi_0}}+{{s\Delta_1}\over{\hbar v}},
\label{eqk}
\end{equation}
where the last term is a SO correction to $k_c^{\rm nr}$ of Eq.~(\ref{eqkc}). We note that $k_c$ is spin and isospin dependent, $k_c=k_{c,\tau,s}$, but to simplify notations we suppress the indices $(\tau,s)$ in what follows.

We note that the curvature and SO corrections to $k_c$ should be applied when calculating energy levels and $z_{k_c,k_t}$ amplitudes of Eq.~(\ref{elnt}). However, the phase factors of the wavefunctions $\Psi^\tau_{k_c,k_t}$ are fixed strictly by the periodicity condition and cannot be changed by the renormalization of $k_c$ nor by minor changes in $R$ and ${\bf K}^\tau$ due to the deformation of graphene when folding it into a nanotube. Finally,  phase factors of the circumferential wavefunctions can be chosen quite generally as
\begin{equation}
e^{i({\bf K^\tau}\cdot{\bf r}_c)}e^{ik_cc}=\exp{\left[i\left(\tau M+n_2{{1+\tau}\over{2}}\right)\varphi\right]}
\label{phase}
\end{equation}
for the lowest circumferencial modes, $m=0$, where $\varphi$ is the azimuth along the circumference ${\bf C}$, and ${\bf r}_c$ is the circumferential component of $\bf r$. In Eq.~(\ref{phase}), the two first terms in the parentheses stem from the dot products $({\bf K^\tau}\cdot{\bf C})$ and the $\nu$ term is canceled by the leading term in $k_c$ of Eq.~(\ref{kg}). This equation will be used in Sec.~\ref{CoulEl} below to calculate Coulomb integrals.

According to Eq.~\eqref{spenergy}, the spin degeneracy of the single-particle levels is lifted by SO coupling that in the absence of a magnetic field opens a gap between the spin $s=\mid\uparrow\rangle$ and $s=\mid\downarrow\rangle$ states of size $\Delta_{SO}(k_t)$ for each of the ${\bf K}^{\tau}$ points. It is defined as
\begin{equation}
\Delta_{SO}(k_t)=\left\vert E_{k_c,k_t}^{+\uparrow}-E_{k_c,k_t}^{+\downarrow}\right\vert=\left\vert E_{k_c,k_t}^{-\downarrow}-E_{k_c,k_t}^{-\uparrow}\right\vert,
\label{SOgap0}
\end{equation}
where $E_{k_c,k_t}$ is defined in Eq.~\eqref{engraph}. For $B=0$, energies $E^{\tau,s}_{k_c,k_t}$  depend only on the product $\tau s$ and therefore coincide at both ${\bf K}^\tau$ points. Because of the axial magnetic field, this degeneracy is lifted and the spectrum in Eq.~(\ref{spenergy}) consists of two Kramers doublets. Using Eq.~(\ref{eqk}) and expanding Eq.~\eqref{spenergy} in the
small parameters $\Delta_1/k_g\ll 1, k_t/k_g\ll1, \Phi_{AB}/(\Phi_0R k_g)\ll 1$, we obtain
\begin{eqnarray}
E_{\pm}^{\tau s}(B)&=&\pm\hbar v |k_g|+\tau s\left(\Delta_0\pm \mbox{sign}\{k_g\}\Delta_1\right)\nonumber\\
&+&s\left(\frac{1}{2}g\mu_BB\pm \tau s~\mbox{sign}\{k_g\}\hbar v\frac{\Phi_{AB}}{R\Phi_0}\right),
\label{enapprox}
\end{eqnarray}
where $\pm$ refers to the conduction and valence bands, respectively, and $\mbox{sign}\{k_g\}\equiv k_g/|k_g|$. We will use Eq.~(\ref{enapprox}) in conjunction with the experimental data for $\Delta_{SO}$ of the electron and hole Kramers doublets of Fig.~4(c) of Ref.~\onlinecite{Kuemmeth}, including their
$B$-dependencies, in order to find $\Delta_0$ and $\Delta_1$. This can only be done with the accuracy to the sign of $k_g$. This sign remains unknown and there are still controversies in the literature encountering different theoretical models, e.g., see discussion in \S 5 of Ref.~\onlinecite{Ando}. In what follows, we accept $k_g<0$.

Because the slopes of the $B$-dependencies in Fig.~4(c) of Ref.~\onlinecite{Kuemmeth} are larger for the lower components of the Kramers doublets, both for electrons and holes, Eq.~(\ref{enapprox}) suggests that
\begin{equation}
\tau s\,\mbox{sign}(k_g)=\pm 1,
\label{staukg}
\end{equation}
for the conduction and valence band, respectively. Then, with $k_g<0$, it immediately follows from Eq.~(\ref{enapprox}) that
\begin{equation}
\Delta_{SO}^e=2(\Delta_0-\Delta_1),\qquad \Delta_{SO}^h=-2(\Delta_0+\Delta_1),
\label{deltas}
\end{equation}
and, with the experimental values $\Delta_{SO}^e=0.37$ meV and $\Delta_{SO}^h=0.21$ meV,\cite{Kuemmeth} we arrive at $\Delta_1= -(\Delta_{SO}^e+\Delta_{SO}^h)/4\approx-0.15$ meV and $\Delta_0= (\Delta_{SO}^e-\Delta_{SO}^h)/4\approx0.04$ meV. These parameter values will be used in all calculations below.

In what follows, we restrict the analysis to the lowest longitudinal mode. While the effect of mode mixing due to Coulomb repulsion is essential in suspended wide-gap nanotubes,\cite{Wunsch,Rontani,Stecher} in coated narrow-gap nanotubes  it is suppressed due to a small effective mass and large $\kappa$.

\subsection{Nanotube Double Quantum Dots - Single electron regime}
\label{DQDsol}
\begin{figure}[t]
\begin{center}	
\includegraphics[width=0.5\textwidth]{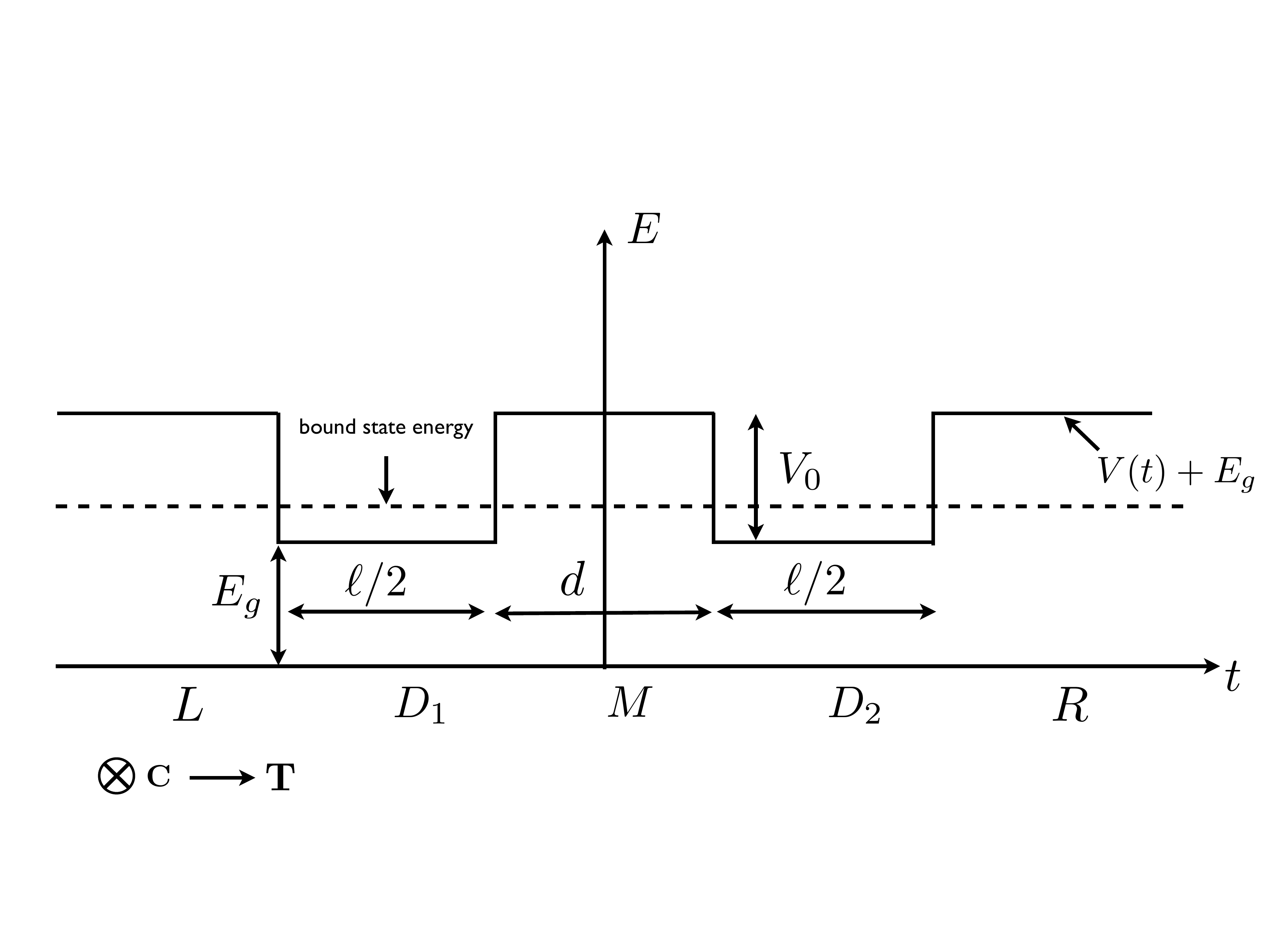}
\caption{ \label{DQD}
Energy diagram of a symmetrical CNT-DQD with electrostatic gates inducing a potential $V(t)$ along the nanotube axis; the origin is chosen in the center of the double dot, and the energy reference point is chosen in the middle of the gap. Two dots $D_1$ and $D_2$ of length $\ell /2$ each are tunnel coupled via a middle barrier $M$ of width $d$. The bound state energy $E$ (shown by dashed line) obeys the criterion $E_g<E<V_0+E_g$.
}
\end{center}
\end{figure}
In this section we outline a model of a CNT-DQD motivated by
the experimental  setup in Ref.~\onlinecite{Churchill}. The
underlying geometry is sketched in Fig.~\ref{DQD}. The
potential along the nanotube is controlled by top gates. For
simplicity, we choose a double-well potential to model the
double dot, similarly to Refs.~\onlinecite{Bulaev} and
\onlinecite{Wunsch},
\begin{equation}
V(t)=\left\{
\begin{array}{rcl}
V_0,&& |t|<d/2,\,\,|t|>(d+\ell)/2\\
0,&& d/2\leq |t|<(d+\ell)/2
\end{array}
\right. \label{pot}
\end{equation}
and solve the eigenvalue problem for the Hamiltonian
\begin{equation}
H(t)=\hbar v(\tau k_c\sigma_1+k_t\sigma_2)+\left(\Delta_0\tau+{1\over2}g\mu_BB\right)s+V(t),
\label{hamV}
\end{equation}
with $k_c$ of Eq.~(\ref{eqk}).
The potential $V(t)$ is considered as step-like on the scale of the Fermi wavelength, $2\pi/k_t$, but smooth on the scale of the inverse Brillouin momentum, $2\pi/|K^\tau|$. Therefore, it does not induce essential ${\bf K}^+{\bf K}^-$-scattering. This is indeed relevant for an electrostatically confined dot.  We calculate the single electron wavefunction for the symmetric geometry shown in Fig.~\ref{DQD} for equal confining potentials of the left and right wells, $V_1=V_2=V_0$. The two QDs are connected by a barrier of width $d$. The generalization to an asymmetric geometry is straightforward and will result in a larger electron wavenumber for the deeper well. The total energy of the  system,
$E=E_1+E_2$, will be considered fixed.
In Eq.~(\ref{hamV}), the term in the second parentheses produces an $(s, \tau)$ dependent level shift but does not influence the wave functions. For the electronic wavefunction $\psi^{\tau}(t)$, defined in different intervals, see Fig.~\ref{DQD}, we use the ansatz
\begin{eqnarray}
\psi^\tau_L(t)&=&Ae^{q_{t}t}\left(\begin{array}{cc}z^\tau _{k_c,-iq_{t}}\\ 1\end{array}\right),\nonumber\\
\psi^\tau_{D_1}(t)&=&Be^{ik_{t}t}\left(\begin{array}{cc}z^\tau_{k_c,k_{t}}\\ 1\end{array}\right)+Ce^{-ik_{t}t}\left(\begin{array}{cc}z^\tau_{k_c,-k_{t}}\\ 1\end{array}\right),\nonumber\\
\psi^\tau_M(t)&=&De^{-q_{t}t}\left(\begin{array}{cc}z^\tau_{k_c,iq_{t}}\\ 1\end{array}\right)+Ee^{q_{t}t}\left(\begin{array}{cc}z^\tau_{k_c,-iq_{t}}\\ 1\end{array}\right),\nonumber\\
\psi^\tau_{D_2}(t)&=&Fe^{ik_{t}t}\left(\begin{array}{cc}z^\tau_{k_c,k_{t}}\\ 1\end{array}\right)+Ge^{-ik_{t}t}\left(\begin{array}{cc}z^\tau_{k_c,-k_{t}}\\ 1\end{array}\right),\nonumber\\
\psi^\tau_R(t)&=&He^{-q_{t}t}\left(\begin{array}{cc}z^\tau_{k_c,iq_{t}}\\ 1\end{array}\right),\nonumber\\
\label{ansatzwave}
\end{eqnarray}
with $z^\tau$ factors defined for electrons and holes, respectively, according to Eq.~(\ref{zfunc}).
In the gate regions, the electron wavenumber is determined as
\begin{equation}
q_{t}=\sqrt{k_c^2-(E_{k_c,k_{t}}-V_0)^2/(\hbar v)^2}.
\label{qt}
\end{equation}
Since the Hamiltonian $H(t)$ of Eq.~(\ref{hamV}) remains invariant under complex conjugation accompanied by $k_t\rightarrow -k_t$, the wavefunctions of Eq.~(\ref{ansatzwave}) can be chosen in such a way that $C=B^*$, $G=F^*$, and $A$, $D$, $E$, and $H$ are real.

Energy levels are found from a transcendental equation that follows from the continuity of the wavefunction $\psi^\tau(t)$ at all potential steps. Since $v_t=\partial H/\partial p_t =v\sigma_2$, this equation maintains the current continuity. All coefficients in Eq.~(\ref{ansatzwave}) can be calculated from the boundary conditions and the normalization condition. For $0<V_0<2E_g$, at least one electron bound state exists in each of the dots, and we will restrict ourselves only with the ground state (GS) mode. In a double dot, it splits into two, bonding and antibonding, with energies $E_b$ and $E_{ab}$, respectively .

\begin{figure}[t!]
\begin{center}	
\begin{minipage}{80mm}
\includegraphics[width=80mm]{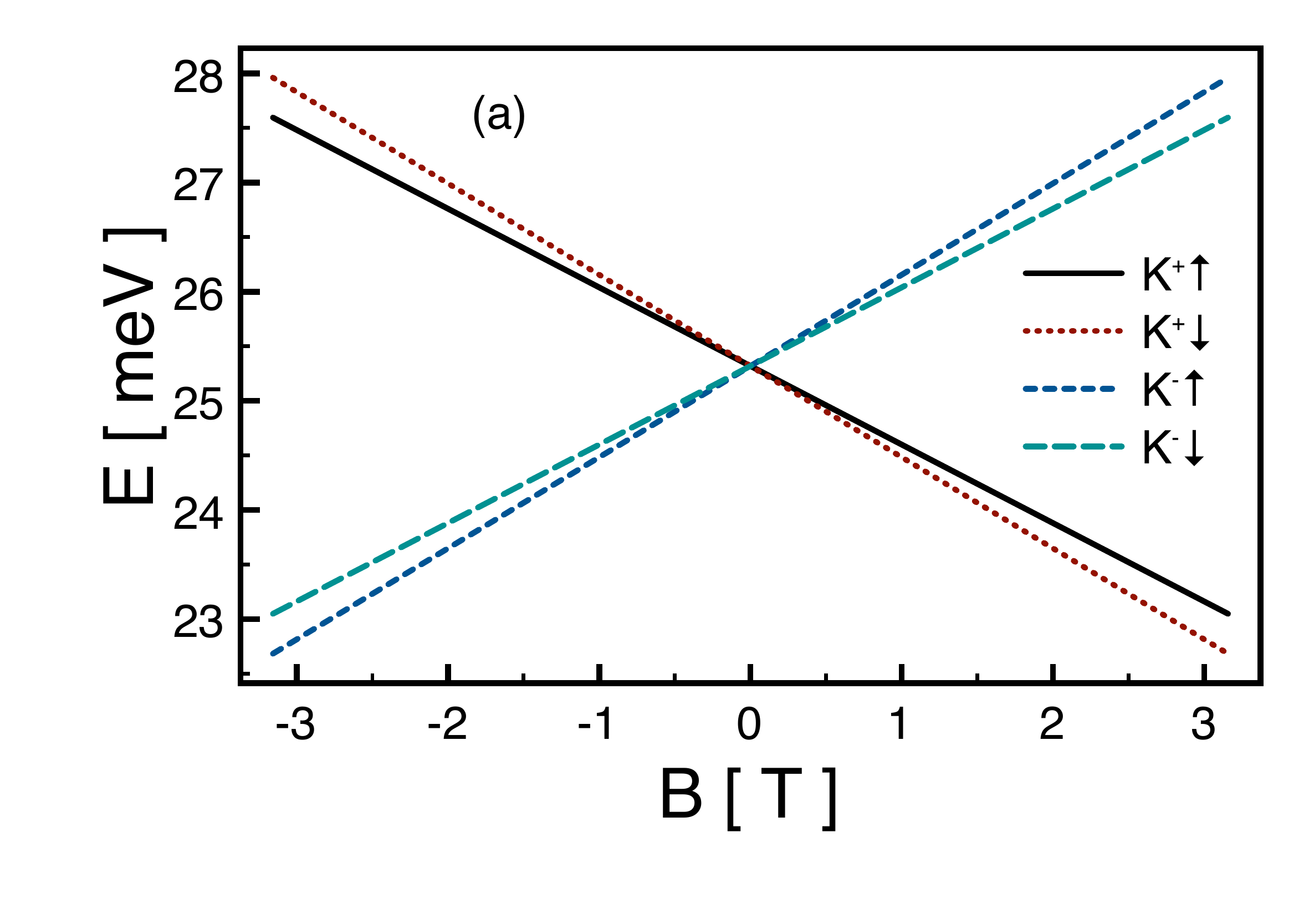}
\end{minipage}
\hspace{10mm}
\begin{minipage}{80mm}
\includegraphics[width=80mm]{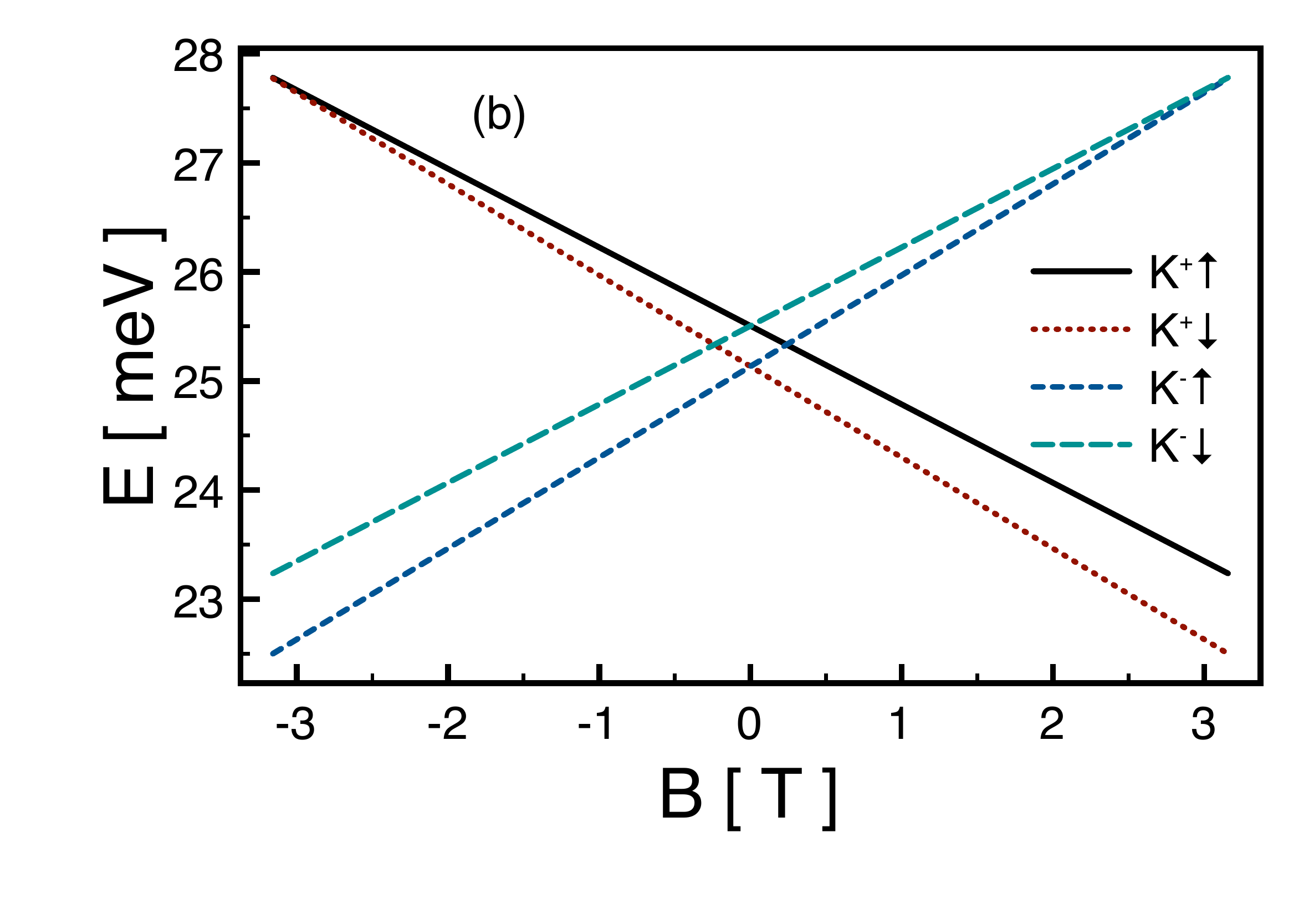}
\end{minipage}
\caption{ \label{spdqd} (color online) Magnetic field dependence of the energy of the bonding
 mode of a single-electron DQD for $k_g=-0.045 $ nm$^{-1}$ and $\ell=200$ nm, $d=120$ nm, $V_0=3.95$ meV. The ground state energy $E_{GS}$ is close to the gap edge, $E_{GS}-E_g=1.4$ meV. (a) SO coupling is absent, $\Delta_{SO}=0$. At $B=0$, the spectrum is four-fold degenerate. (b)  SO split energy spectrum. The zero field splitting between Kramers doublets is $\Delta_{SO}=0.37$ meV (with $\Delta_1=-0.15$ meV and $\Delta_0=0.04$ meV), see text for details.}
\end{center}
\end{figure}

The energy spectrum of the bonding mode of the Hamiltonian
$H(t)$ of Eq.~(\ref{hamV}) is plotted in Fig.~\ref{spdqd}(a) as
a function of an axial magnetic field in the absence of SO
coupling. Having in mind narrow gap nanotubes,\cite{Churchill}
we choose $k_g=-0.045 \mbox{ nm}^{-1}$, which produces a gap of
about $2E_g\approx 46$ meV that is eight times less than
estimated according to Eq.~\eqref{kg} for $\nu=1$
semiconducting tubes with $R\sim 1\mbox{ nm}$. In the absence
of SO coupling, the $B=0$ state is four-fold degenerate in $s$
and $\tau$, and its splitting by magnetic field $B$ is shown in
Fig.~\ref{spdqd}(a). Fig.~\ref{spdqd}(b) demonstrates the
splitting of the $B=0$ state by SO coupling into a doublet of
two pairs of Kramers conjugate states, ($|K^+_\uparrow\rangle,
|K^-_\downarrow\rangle$) and
$(|K^+_\downarrow\rangle,|K^-_\uparrow\rangle)$. The zero-field
splitting is $\Delta_{SO}^e\approx 0.37$ meV, according to the
data of Ref.~\onlinecite{Kuemmeth} and in a reasonable
agreement with the data of Ref.~\onlinecite{Churchill}. The
$B$-dependence in Fig.~\ref{spdqd}(b) reproduces correctly the
patterns observed experimentally in Refs.~\onlinecite{Kuemmeth}
and \onlinecite{Churchill}. The effective $g$-factor, $\partial
\Delta E/\partial (\mu_B B_{||})$, of the upper (lower) doublet
is smaller (larger) due to the coupling of real spin to
isospin.

The upper branch crossing in Fig.~\ref{spdqd}(b) persists for ${\bf B}\parallel{\bf T}$, but turns into an avoided crossing when $\bf B$ acquires a perpendicular component.\cite{Churchill} Spin relaxation near this crossing was investigated in Refs.~\onlinecite{Churchill}, \onlinecite{Bulaev}, and \onlinecite{Rudner}. The lower branch crossing, at much lesser fields $B$, turns into an avoided crossing by ${\bf K}^+{\bf K}^-$ scattering with large momentum transfer, $\Delta_{K^+K^-}$. In high quality nanotubes of Refs.~\onlinecite{Kuemmeth} and \onlinecite{Churchill}, where both $\Delta_{SO}$ and $\Delta_{K^+K^-}$ were observed experimentally, $\Delta_{K^+K^-}$ was nearly one order of magnitude less than $\Delta_{SO}$; in what follows, we disregard $\Delta_{K^+K^-}$. Its origin is not well understood for now, and it is usually attributed to electron scattering by defects.\cite{Bulaev}

\begin{figure}[t!]
\begin{center}
\begin{minipage}{80mm}	
\includegraphics[width=80mm]{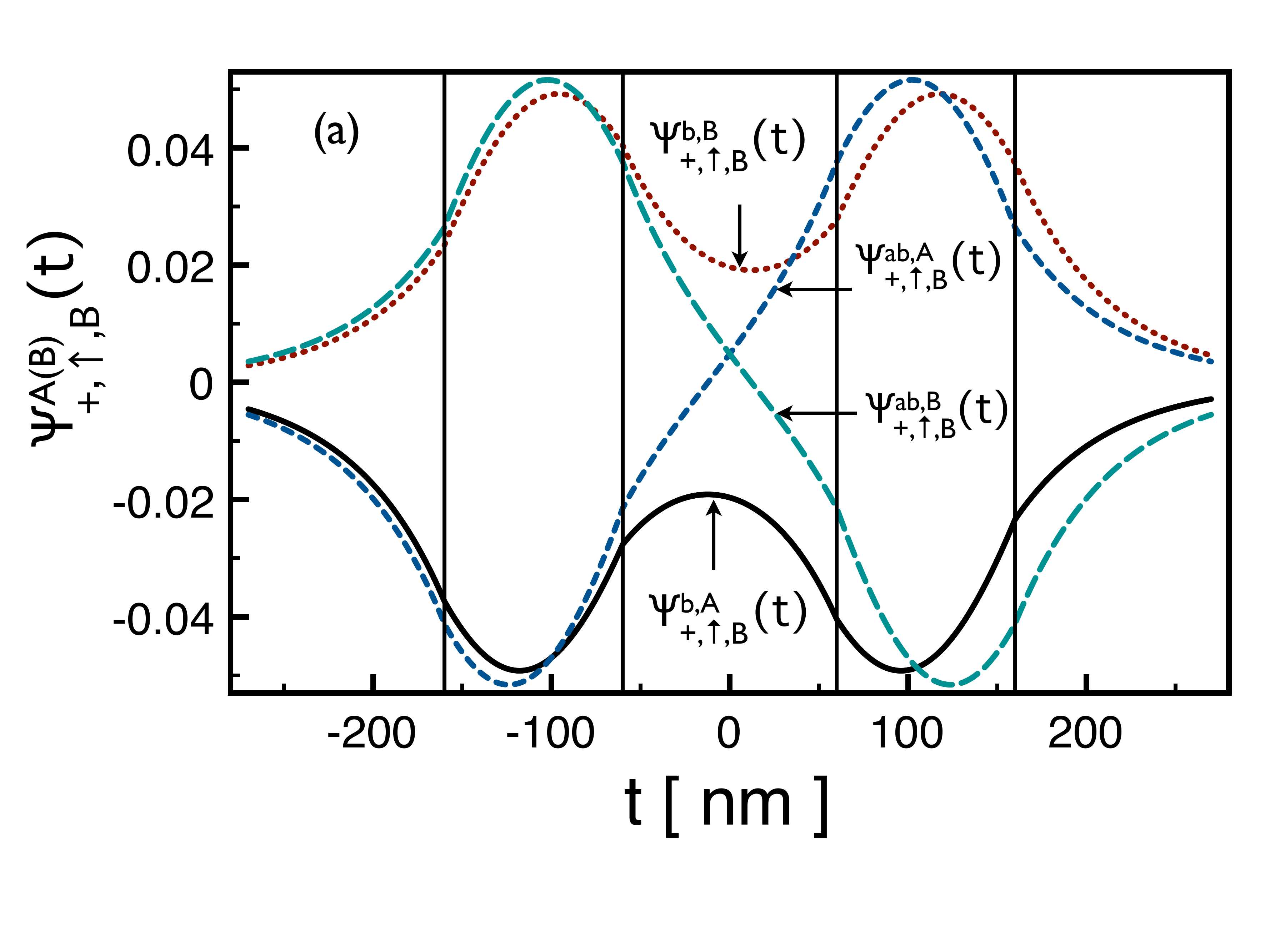}
\end{minipage}
\hspace{10mm}
\begin{minipage}{80mm}
\includegraphics[width=80mm]{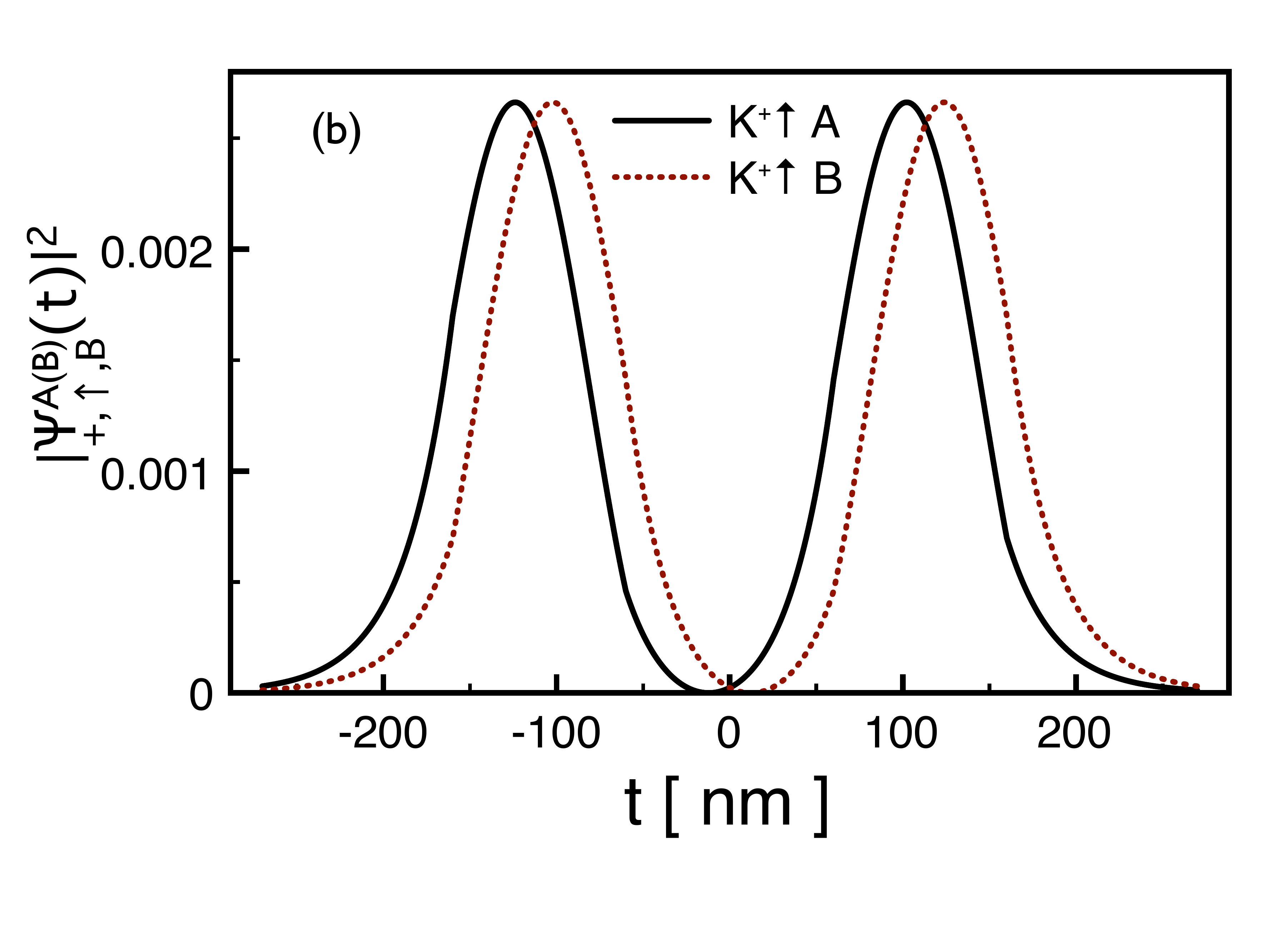}
\end{minipage}
\caption{ \label{wavefun} (color online) Single particle wavefunctions $\psi^{A(B)}_{+,\uparrow,B}(t)$ of a DQD
 for the same parameter values as in Fig.~\ref{spdqd} and $B=1$ T. Vertical lines sketch edges of both dots. (a) Real parts of $\psi^{A(B)}_{+,\uparrow,B}(t)$ on both $A(B)$ sublattices are shown for bonding and antibonding modes, denoted by the superscripts $a$ and $ab$, respectively. At $t=0$ the former modes retain a considerable amplitude, whereas the latter vanish at the origin. Note that at each point the $A$ and $B$ components have nearly the same absolute values but opposite signs.
(b) Electron density distribution in the bonding state; a non-vanishing density at $t=0$ is visible as well as a $t\rightarrow-t$ asymmetry and difference between $A$ and $B$ densities.}
\end{center}
\end{figure}

Next, we discuss the longitudinal part of the wavefunction $\psi^\tau(t)$ and begin with exact symmetries of wavefunctions. Substituting Eq.~(\ref{eqk}) into Eq.~(\ref{hamV}), one arrives at a Hamiltonian $H_{\tau,s,B}(t)$. It includes the parameters $(\tau,s,B)$ only as time-inversion symmetric
products $\tau B$ and $\tau s$. Therefore, $H_{-\tau,-s,-B}(t)=H_{\tau,s,B}(t)$, and wavefunctions can be chosen in such a way that they possess the same symmetry
\begin{equation}
\psi^{A(B)}_{\tau,s,B}(t)=\psi^{A(B)}_{-\tau,-s,-B}(t),
\label{psi1}
\end{equation}
for an arbitrary potential $V(t)$. The superscripts $A$ and $B$ indicate sublattice indices. For symmetric dots, $V(t)=V(-t)$, one more relation holds. Performing a canonical transformation of $H_{\tau,s,B}(t)$ with a matrix $\sigma_1$, one notices that the sign change of the term $k_t\sigma_2$ can be compensated by a $t\rightarrow-t$ transformation. Because the $\sigma_1$ transformation transposes sublattices, $\sigma_1\left(\begin{array}{c}\psi^A\\\psi^B\end{array}\right)=\left(\begin{array}{c}\psi^B\\\psi^A\end{array}\right)$, wavefunctions obey the further relations
\begin{equation}
\psi^{A(B)}_{\tau,s,B}(t)=\pm\psi^{B(A)}_{\tau,s,B}(-t).
\label{psi2}
\end{equation}
Here and in what follows we choose wavefunctions in such a way
that they are real in the classically forbidden regions, which
is always possible because the $z$-factors of Eq.~(\ref{zfunc})
are real there. We note that in the left and right hand sides
of Eq.~(\ref{psi2}) the $A(B)$ sublattices are interchanged.
This reflects an absence of the microscopic $t\rightarrow - t$
symmetry in chiral nanotubes. Therefore, it is also absent in
``symmetric" DQDs despite the macroscopic symmetry of the
confining potential. Manifestations of this asymmetry are
discussed next. Numerical data for the wavefunctions are
presented in Fig.~\ref{wavefun}. Due to the exact symmetries of
Eqs.~(\ref{psi1}) and (\ref{psi2}), it is enough to display
only a few curves demonstrating the basic regularities. Real
parts of the wavefunctions of the bonding and antibonding modes
are shown in Fig.~\ref{wavefun}(a). Their different behavior
inside the tunnel barrier is distinctly seen. Also, $A$ and $B$
components of wavefunctions are nearly symmetric and
antisymmetric for the bonding and antibonding modes,
respectively. Technically, the asymmetry arises due to the
admixture of the valence band wavefunctions and is small
because the GS binding energy, $E_{GS}-E_g\approx1.4$ meV, is
small compared to the gap, $E_g\approx 23$ meV (or, what is
essentially the same, $k_t,q_t\ll |k_g|$). This asymmetry
increases with $(E_{GS}-E_g)$ and can become of the order of
unity for $|E_{GS}-E_g|\sim E_g$.\cite{Rudner} The opposite
signs of the functions $\psi^A(t)$ and $\psi^B(t)$ originate
from $k_g<0$. Fig.~\ref{wavefun}(b) displays small differences
in the electron densities on both sublattices and their
asymmetries that have the same origin as in
Fig.~\ref{wavefun}(a).

Within the GS approximation, the appearance of bonding (lower) and antibonding (upper) tunnel components, denoted as
 $\psi^b$ and $\psi^{ab}$, respectively, see Fig.~\ref{wavefun}, motivates a treatment of the orbital degrees of freedom of a DQD in terms of an effective
two level system. Its  eigenstates may be obtained by hybridization of the electron states localized in $D_1$ or $D_2$.
Using the wavefunctions of Eq.~\eqref{ansatzwave} and notations of
Eq.~\eqref{psi2}, we define the orbital basis states for certain spin and isospin (i.e., $|LK^\tau_s \rangle$ and $|RK^\tau_s\rangle$) in the left and right halfspaces, respectively,  as
\begin{eqnarray}
|LK^\tau_s \rangle&=&\frac{1}{\sqrt{2}}\left[\psi_{\tau,s,B}^b(t)+\psi_{\tau,s,B}^{ab}(t)\right],\nonumber\\
|RK^\tau_s \rangle&=&\frac{1}{\sqrt{2}}\left[\psi_{\tau,s,B}^b(t)-\psi_{\tau,s,B}^{ab}(t)\right],
\label{LRwave}
\end{eqnarray}
with associated energies $E_{\tau s}(B)$; these energies include also the second term of Eq.~(\ref{hamV}). Each of these functions is a two-spinor in sublattice space defined by its $A(B)$ components, and $K^\tau_s$ indicates that they should be chosen with $k_c$ of Eq.~(\ref{eqk}) calculated for proper values of $(\tau,s)$. This parametric dependence of the orbital functions on $s$ and $\tau$ stems from SO coupling and will be essential for the classification of quantum states, see Sec.~\ref{sec:classification} below.

The connection between these basis states via tunneling is naturally established by the overlap of the wavefunctions in the interval $|t|\leq d/2$.
The resulting spin and isospin conserving Hamiltonian expressed in this basis is
\begin{equation}
H_{RL}
=\left(\begin{array}{cc}E_{k_c,k_t}^{\tau sR}-\varepsilon&\eta_{\tau,s}\\\eta_{\tau,s}&E_{k_c,k_t}^{\tau sL}+\varepsilon\end{array}\right),
\label{TLSham}
\end{equation}
where $E_{k_c,k_t}^{\tau sL}=E_{k_c,k_t}^{\tau sR}=\frac{1}{2}\left(E_{b}^{\tau s}+E_{ab}^{\tau s}\right)$ are the single particle energies
and  $\varepsilon$ is the detuning between the left and right dot.  The  connection between left and right halfspaces, established by tunneling,
produces bonding and antibonding states of Eq.~\eqref{hamV}. The tunnel matrix element $\eta_{\tau,s}$ is related to their energy difference as
\begin{equation}
2\eta_{\tau,s}= E^{\tau s}
_{b}-E^{\tau s}_{ab}.
\label{tuneq}
\end{equation}
Equivalently, the tunneling part $H_T$ of the Hamiltonian of Eq.~(\ref{TLSham}) can be rewritten in terms of the orbital functions $|L(R)K^\tau_s\rangle$ in a form that is more convenient for further calculations
\begin{equation}
H_T=\sum_{\tau=\pm,s=\uparrow,\downarrow}\eta_{\tau,s}\left[
|LK^\tau_{s} \rangle^{A(B)}\langle RK^\tau_{s}|^{A(B)}+ h.c.\right] .
\label{Htun}
\end{equation}

\begin{figure}[t!]
\begin{center}	
\begin{minipage}{80mm}
\includegraphics[width=80mm]{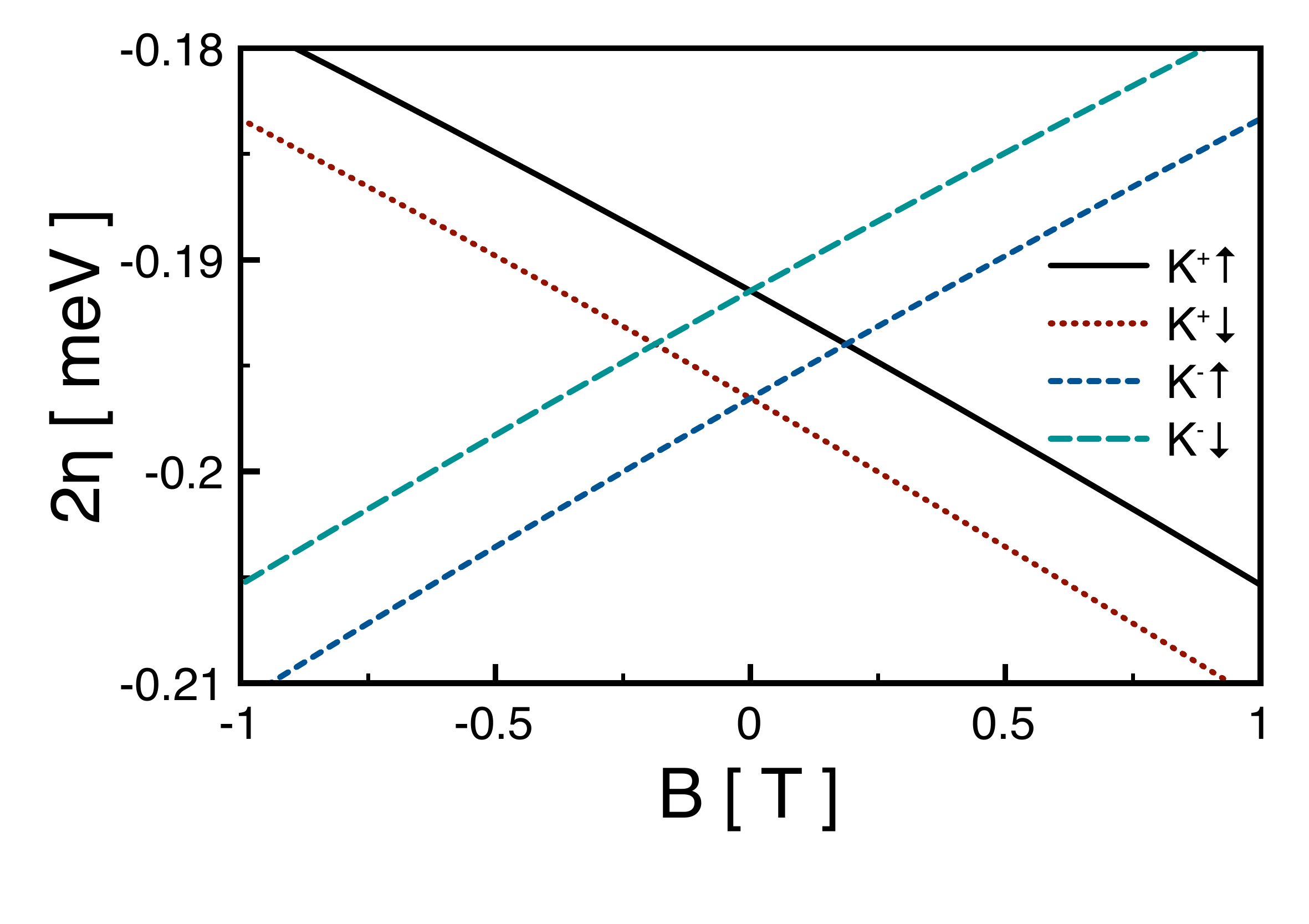}
\end{minipage}
\caption{ \label{tunnel} (color online) Tunneling matrix elements  $\eta_{\tau,s}(B)$ plotted versus magnetic field. Their absolute values are larger for the lower  Kramers doublet $(K^+_{\downarrow},K^-_{\uparrow})$. For $B=0$, they depend on spin and isospin through the product $s\tau$. The zero field difference in the tunneling rate originates from $\Delta_1\neq 0$ and vanishes for $\Delta_1\to 0$. Parameter values are the same as in Fig.~\ref{spdqd}(b).}
\end{center}
\end{figure}

The matrix elements $\eta_{\tau,s}$ are plotted in Fig.~\ref{tunnel} as a function of $B$ for the parameter values used in Fig.~\ref{spdqd}(b). The zero field difference in the tunneling matrix elements stems from the $\Delta_1$ term in Eq.~(\ref{eqk}). They are nearly linear functions of $B$ in the whole region of magnetic field and their absolute values are larger for the lower Kramers doublet. This fact is also reflected in the following inequality for the  wavenumbers in the classically forbidden regions, $q_t(K^+_\uparrow)>q_t(K^+_\downarrow)$, cf.~Eq.~\eqref{qt}, that holds for our set of parameter values. Since the Hamiltonian $H(t)$ of Eq.~\eqref{hamV} is symmetric with respect to time inversion, $\eta_{\tau,s}(B)=\eta_{-\tau, -s}(-B).$

We note that a change of the sign $k_g$, to $k_g>0$, with the values of $\Delta_0$ and $\Delta_1$ recalculated properly to keep the right slopes in Fig.~\ref{spdqd}(b) ($\Delta_0=-0.04$ meV and $\Delta_1=-0.15$ meV), keeps Fig.~\ref{tunnel} practically unchanged with the upper doublet having a lesser tunneling rate (absolute value of $\eta$),
since the sign of $\Delta_1$ remains intact. Therefore, these data do not constrain the sign of $k_g$ either. Such a behavior of the tunneling rates seems counterintuitive, but it stems from the fact that the rates  are controlled by the sign of $\Delta_1$ while the nature of the upper and lower Kramers doublets is insensitive to it.

\begin{figure}
\begin{minipage}[t!]{80mm}
\begin{center}	
\includegraphics[width=80mm]{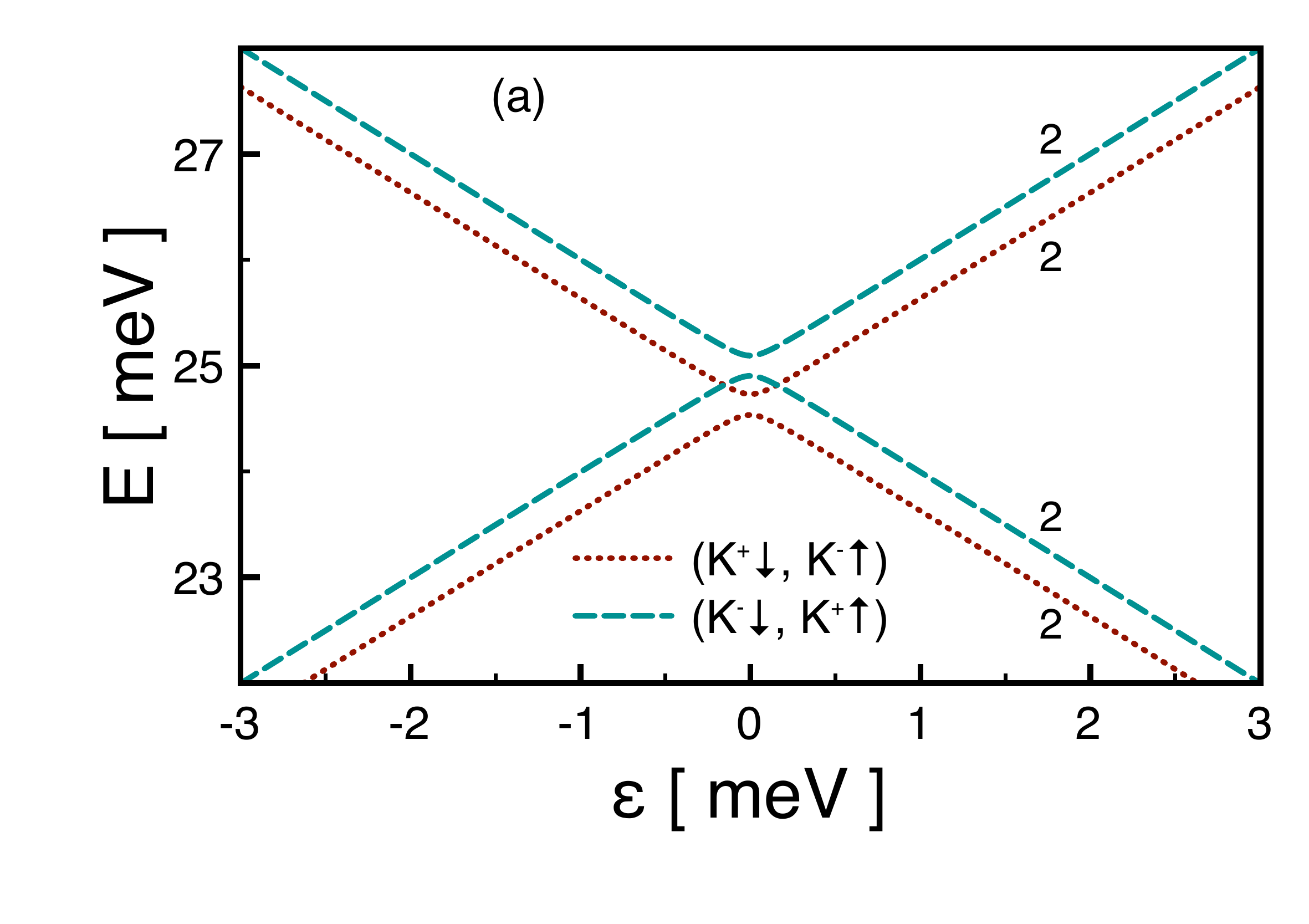}
\end{center}
\end{minipage}
\begin{minipage}[t!]{80mm}
\begin{center}	
\includegraphics[width=80mm]{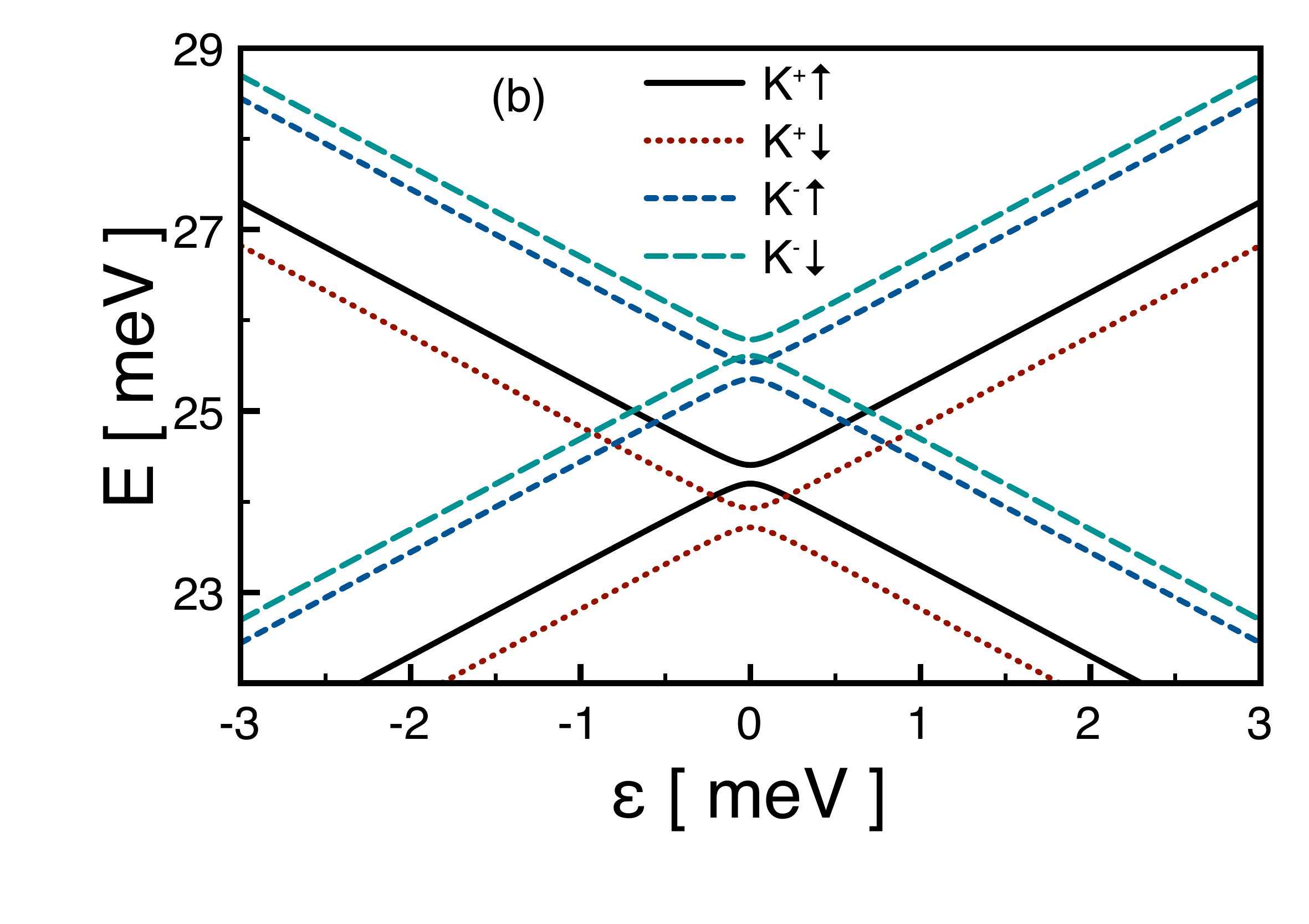}
\end{center}
\end{minipage}
\caption{ \label{0110a} (color online) Detuning dependence of the energy spectrum of a single-electron DQD. Tunnel matrix elements $\eta_{\tau,s}$ from Fig.~\ref{tunnel} are used for the respective states, other parameters as in Fig.~\ref{spdqd}. (a) Both Kramers doublets (identified by color) split into bonding and antibonding modes.
The numbers denote energy level multiplicity. (b) Level splitting by magnetic field $B=1$ T. Because of the dominating effect of the Aharonov-Bohm flux, $K^-$ components with both $s=\uparrow,\downarrow$ move upwards. The magnitudes of the avoided crossings at $\varepsilon=0$, seen both in (a) and (b), are controlled by $\eta_{\tau,s}$.}
\end{figure}

Fig.~\ref{0110a} shows the effect of detuning $\varepsilon$ on the energy spectrum of the Hamiltonian Eq.~\eqref{TLSham}. In our calculations, we have consistently used the  $\tau,s,$ and $B$ dependencies of the matrix elements $\eta_{\tau,s}(B)$ derived from Eq.~\eqref{tuneq}.
Cyan (dash-dotted) and red (dotted) lines in Fig.~\ref{0110a}(a) correspond to the upper and lower Kramers doublets. In the absence of a magnetic field
all states are two-fold Kramers degenerate. For $B\neq0$ this degeneracy is lifted; the spectrum for $B=1\mbox{ T}$ is shown in Fig.~\ref{0110a}(b). Energies corresponding  to various bonding (antibonding) and spin (isospin) states are shown in the same colors and line patterns as in Fig.~\ref{spdqd}. The spectrum splitting in a magnetic field originates from the Aharonov-Bohm flux $\Phi_{AB}$ of Eq.~(\ref{eqk}) (that can be ascribed to an orbital magnetic moment $\mu_{\rm orb}$) and the Zeeman spin splitting described by the next-to-last term in Eq.~(\ref{hamV}). Because $\mu_{\rm orb}\sim10 \mu_B$,\cite{Minot,Kuemmeth,Churchill} the first contribution dominates and both spin components of the $K^-$ state move up in energy with increasing $B$. The asymmetry in the level splittings inside the upper and lower quadruplets in the large $B$ region originates from the interference of the SO and Zeeman splittings in each of the single dots.\cite{Kuemmeth,Churchill}

\section{Two electron regime}
\label{twopart}
Charge states in DQD systems are usually described in terms of stability diagrams.\cite{vanderWiel, Hanson, Churchill} Either from Coulomb blockade peaks in transport measurements or from charge sensing probes, the number of electrons $n_L$ and $n_R$ in the respective dot is monitored. In the following we consider the two electron regime. There are two possible physical realizations of it depending on the adjustment of the confinement potentials for the left and right dot and the tunneling barrier between them. First, two electrons are confined to a single dot, here the right dot, denoted the $(02)$ configuration. Otherwise, both electrons are confined in different dots, which is referred to as the $(11)$ configuration. Experimentally, it is possible to drive the DQD between the two regimes by applied gate voltages if both dots are connected by a tunneling barrier.

\subsection{$(02)$ Configuration}
\label{sec:classification}

For a two-electron system without SO coupling, a classification of two-electron states in terms of singlet and triplet states in real spin is exact due to the Pauli principle.\cite{QuantMech} Here, when constructing two-electron states, we use the Hilbert space spanned by the lowest energy orbitals in each dot. Because the spin and orbital degrees of freedom are coupled, a classification in terms of spin singlet and triplets is no longer applicable. Technically, this means that a full basis of two-particles states in the space of functions respecting the Pauli exclusion principle cannot be constructed in terms of products of spin singlets (triplets) and linear combinations of products of the orbital functions $|L(R)K^\tau_s\rangle$ of Eq.~(\ref{LRwave}). Spin singlets and triplets become coupled, and using such generalized basis states we arrive at the results that are quite general and, in particular, can be applied in the vicinity of the point where $k_c$ vanishes due to the cancellation of different terms in Eq.~(\ref{eqk}); the latter regime has been reported recently for a single dot.\cite{Jhang} However, the regime where the gap between the conduction and valence bands closes completely (see Ref.~\onlinecite{Jhang}) is not addressed here, since it does not allow for electrostatic confinement of electrons or holes.

With two electrons bound to the right dot, there are six linearly independent antisymmetric basis functions. Two of them, with both spins either up or down, can be considered as components $T_{\pm1}$ of a spin triplet. They are
\begin{eqnarray}
|\Phi^{02}_s\rangle={{1}\over{\sqrt{2}}}\left(|RK^+_ss\rangle_1|RK^-_ss\rangle_2-1\rightleftharpoons2\right),
\label{ST1}
\end{eqnarray}
with $|\Phi^{02}_1\rangle$ for $s=\uparrow$ and $|\Phi^{02}_2\rangle$ for $s=\downarrow$, and the symbol $1\rightleftharpoons2$ indicates electron transposition. Here $|RK^\pm_ss\rangle$ are products of orbital functions $|RK^\pm_s\rangle$ and their spin counterparts $|s\rangle$.

Four different functions, all with opposite spins, are spin-isospin coupled. Within the first pair of states, both electrons reside in the same $K^\tau$ point, but in such a way that in the $B=0$ limit one of the electrons belongs to the upper and the second to the lower Kramers doublet of Fig.~\ref{spdqd}
\begin{eqnarray}
|\Phi^{02}_\tau\rangle={{1}\over{\sqrt{2}}}\left(|RK^\tau_\uparrow\uparrow\rangle_1|RK^\tau_\downarrow\downarrow\rangle_2-1\rightleftharpoons2\right),
\label{calT}
\end{eqnarray}
with $|\Phi^{02}_3\rangle$ for $\tau=+$ and $|\Phi^{02}_4\rangle$ for $\tau=-$. In the second pair of states, for $B=0$, both electrons populate the upper  and lower Kramers doublets of Fig.~\ref{spdqd}
\begin{eqnarray}
|\Phi^{02}_{u}\rangle&=&{{1}\over{\sqrt{2}}}\left(|RK^+_\uparrow\uparrow\rangle_1|RK^-_\downarrow\downarrow\rangle_2-1\rightleftharpoons2\right),\nonumber\\
|\Phi_{l}^{02}\rangle&=&{{1}\over{\sqrt{2}}}\left(|RK^-_\uparrow\uparrow\rangle_1|RK^+_\downarrow\downarrow\rangle_2-1\rightleftharpoons2\right).
\label{calK}
\end{eqnarray}
We designate them as $|\Phi^{02}_5\rangle$ and $|\Phi^{02}_6\rangle$, respectively.

Equations (\ref{ST1})-(\ref{calK}) demonstrate the profound effect of SO coupling on the symmetry of the $(02)$ multiplets: Since spin and isospin are coupled, those states cannot be represented in terms of  spin singlet and triplets. The energy spectrum is SO split. In the absence of a magnetic field, $B=0$,  the state $\mid\Phi_5\rangle$ ($\mid\Phi_6\rangle$) has the highest (lowest) energy because both electrons populate the upper (lower) Kramers doublet, while the four other states are mutually degenerate because one of the electrons populates the upper and the second the lower state.

\subsection{Coulomb  Matrix Elements}
\label{CoulEl}
We choose the Coulomb potential as
\begin{equation}
V_C=\frac{e^2}{4\pi\epsilon_0\kappa}\frac{1}{\sqrt{a_z^2+(t_1-t_2)^2+(2R)^2\sin^2[(\varphi_1-\varphi_2)/2]}},
\label{CP}
\end{equation}
with $\kappa$ as an effective dielectric constant. The cut-off term $a^2_z\approx(2a_B)^2$, with $a_B$ for the Bohr radius, accounts for the size of $2p_z$ functions.\cite{Reinhold} Such a cut-off is convenient in numerical calculations but has no essential effect on
the final results because Coulomb matrix elements converge in two dimensions at small electron separations. In what follows, we calculate matrix elements of $V_C$ for both the $(02)$ and $(11)$ configurations. Similarly, while taking into account consistently the SO corrections to functions $|K^\tau_s\rangle$ in both dots, we keep only spin-diagonal terms when calculating the Coulomb matrix elements and therefore exclude spin nonconserving processes. This approximation is motivated by our focus on dots with small radius $R$ and narrow gap $E_g\approx\hbar vk_c$; indeed, the diagonal SO corrections are large in inverse $k_c$ while nondiagonal SO terms are suppressed for small $R$ by strong orbital quantization in the circular direction.\cite{AndoSO} In exchange matrix elements for the $(02)$ configuration, a selection rule for $\tau$ appears from the fact that $V_C$ depends on $(\varphi_1,\varphi_2)$ only through their difference. Upon using notations $|\tau s\rangle$ for kets including products of the orbital functions $|K^\tau_s\rangle$ and the corresponding spin functions $|s\rangle$, we take advantage of Eq.~(\ref{phase}) and find
\begin{eqnarray}
&&\langle {\tau_1}{s_1}(1),{\tau_2}{s_2}(2)|V_C(1,2)|{\tau_3}{s_3}(2),{\tau_4}{s_4}(1)\rangle\nonumber\\
&\propto&\int_0^{2\pi}d\varphi \exp{[-i(M+n_2/2)(\tau_1+\tau_2-\tau_3-\tau_4)\varphi]},\nonumber\\
\label{SR}
\end{eqnarray}
where $\varphi=(\varphi_1+\varphi_2)/2$. Expressing $M$ in terms of the chiral indices $(n_1,n_2)$ results in $(M+n_2/2)=n_1/3+n_2/6-\nu/3$.  Hence, inside the irreducible wedge of the Bravais lattice where $0\leq\theta<\pi/6$ and $n_1,n_2\geq0$, it is always true that $(M+n_2/2)\neq0$, and therefore
\begin{equation}
 \tau_1+\tau_2=\tau_3+\tau_4.
\label{tauCon}
\end{equation}
Comparing Eq.~(\ref{tauCon}) with the spin selection rule $s_1=s_4$, $s_2=s_3$ (in the leading approximation in SO) underscores a fundamental difference between the spin and isospin, which is an orbital quantum number.

Under these assumptions, $V_C$  of Eq.~(\ref{CP}) is represented in the basis of $\Phi^{02}$-functions of Eqs.~(\ref{ST1})-(\ref{calK}) as
\begin{equation}
{\hat V}_C^{02}=\left(\begin{array}{cccccc}
U_{02}^\uparrow-J^\uparrow_{02}&0&0&0&0&0\\
0&U^\downarrow_{02}-J_{02}^\downarrow&0&0&0&0\\
0&0&U_{02}^+&0&0&0\\
0&0&0&U_{02}^-&0&0\\
0&0&0&0&U_{02}^u&J_{02}\\
0&0&0&0&J_{02}&U_{02}^\ell
\end{array}\right).
\label{VCmat}
\end{equation}
It includes six Coulomb matrix elements $U_{02}$ and three exchange matrix elements $J_{02}$. The latter ones include terms that are nondiagonal in $\tau$ but obey the selection rule of Eq.~(\ref{tauCon}); one can show that the nondiagonal matrix elements $J_{02}$ are real. The absence of isospin indices in $U_{02}^\sigma$ and $J_{02}^\sigma$ indicates that electrons belong to different valleys. Similarly, the absence of spin indices in $U_{02}^\tau$ indicates that electrons possess opposite spins.

All Coulomb terms $U_{02}$ have a universal form in the framework of our model and do not depend on the chirality of the nanotube. As distinct from them, exchange integrals $J_{02}$ depend on chirality and, what is even more important, include products of orbital functions $|RK^\tau_s\rangle$ with different values of $\tau$. Therefore, they require large momentum transfer of $4\pi/3a$, which is accompanied by fast oscillating factors in the integrands. The calculation of such integrals cannot be performed using envelope functions of Eq.~(\ref{elnt}) and requires including
microscopic Bloch functions of graphene and short range interaction potentials. This is outside the framework of our model, and since such integrals are small, we disregard them in what follows.

\begin{figure}[ht!]
\begin{center}	
\includegraphics[width=80mm]{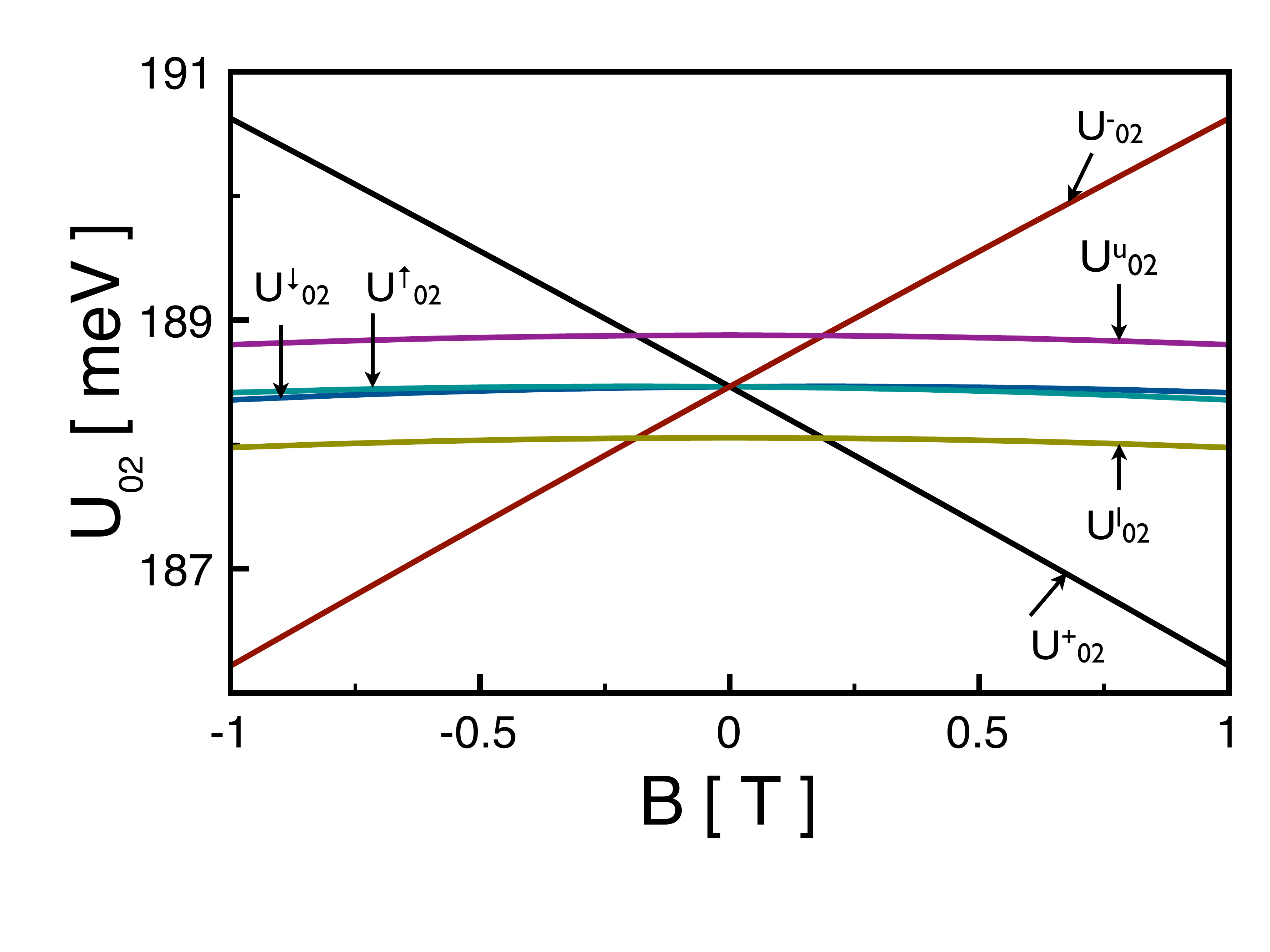}
\caption{ \label{U02ofB}(color online) Magnetic field dependence of the Coulomb matrix elements
$U_{02}$ in a single QD with dielectric constant $\kappa=1$. Parameters are the same as in Fig.~\ref{spdqd}(b). The $B$ dependence is strong for isospin polarized states $|\Phi^{02}_3\rangle$ and $|\Phi^{02}_4\rangle$, with the opposite signs of the slope for $\tau=\pm$. For isospin unpolarized states the $B$-dependences are weak.
The energies $U_{02}^\uparrow$ and $U_{02}^\downarrow$ nearly coincide. For $B=0$, the largest difference in $U_{02}$ energies is achieved for the states with both electrons belonging to the lower or upper Kramers doublet.
The matrix elements $U_{02}$ strongly influence  the position of the $(11)$-$(02)$ degeneracy point in Figs.~\ref{detB0} and \ref{detfinB} below.}
\end{center}
\end{figure}

The six different Coulomb integrals $U_{02}$ and their magnetic field dependence are shown in Fig.~\ref{U02ofB} for $\kappa=1$. As distinct from the single electron levels of Fig.~\ref{spdqd}, where the $B$ dependence originated from the orbital and spin magnetic moments, the $B$ dependence of $U_{02}$ integrals is determined by the $B$ dependence of the circumferencial wavenumber $k_c$ of Eq.~(\ref{eqk}) only. Remarkably, this dependence is much stronger for isospin polarized states $\mid \Phi^{02}_3\rangle$ and $\mid \Phi^{02}_4\rangle$ than for the other four states; for the latter ones, the $B$-dependencies are nearly identical. We attribute this behavior to the competition of the two largest terms in $k_c$, namely the first and second one of Eq.~(\ref{eqk}), which therefore does not rely on SO coupling. In $\mid \Phi^{02}_3\rangle$ and $\mid \Phi^{02}_4\rangle$ both electrons have the same isospin $\tau$, hence the same $B$-dependences of $k_c$ add, while in all other functions the electrons have opposite signs of $\tau$ and the $B$-dependences subtract. We note that in the absence of SO coupling, Coulomb integrals for $\mid \Phi^{02}_1\rangle,\mid \Phi^{02}_2\rangle, \mid \Phi^{02}_5\rangle, \mid \Phi^{02}_6\rangle$ coincide for all magnetic fields. For nanotubes coated by an insulator, as in experiments by Churchill et al.\cite{Churchill,Churchill2}, the Coulomb repulsion is reduced by a factor of $\kappa\approx10$. It should be noted that the metallic gates used in the experimental setups also strongly screen the Coulomb interaction, and in particular they cut-off the long range part of it. Therefore, the absolute numbers for the Coulomb matrix element that we find here are subject to changes depending on the experimental details, however the general trends based on the symmetry of the two-particle wavefunctions remains.
\subsection{Energy spectrum}
\label{sec:EnSp}

\begin{figure}
\begin{minipage}[t!]{80mm}
\begin{center}	
\includegraphics[width=80mm]{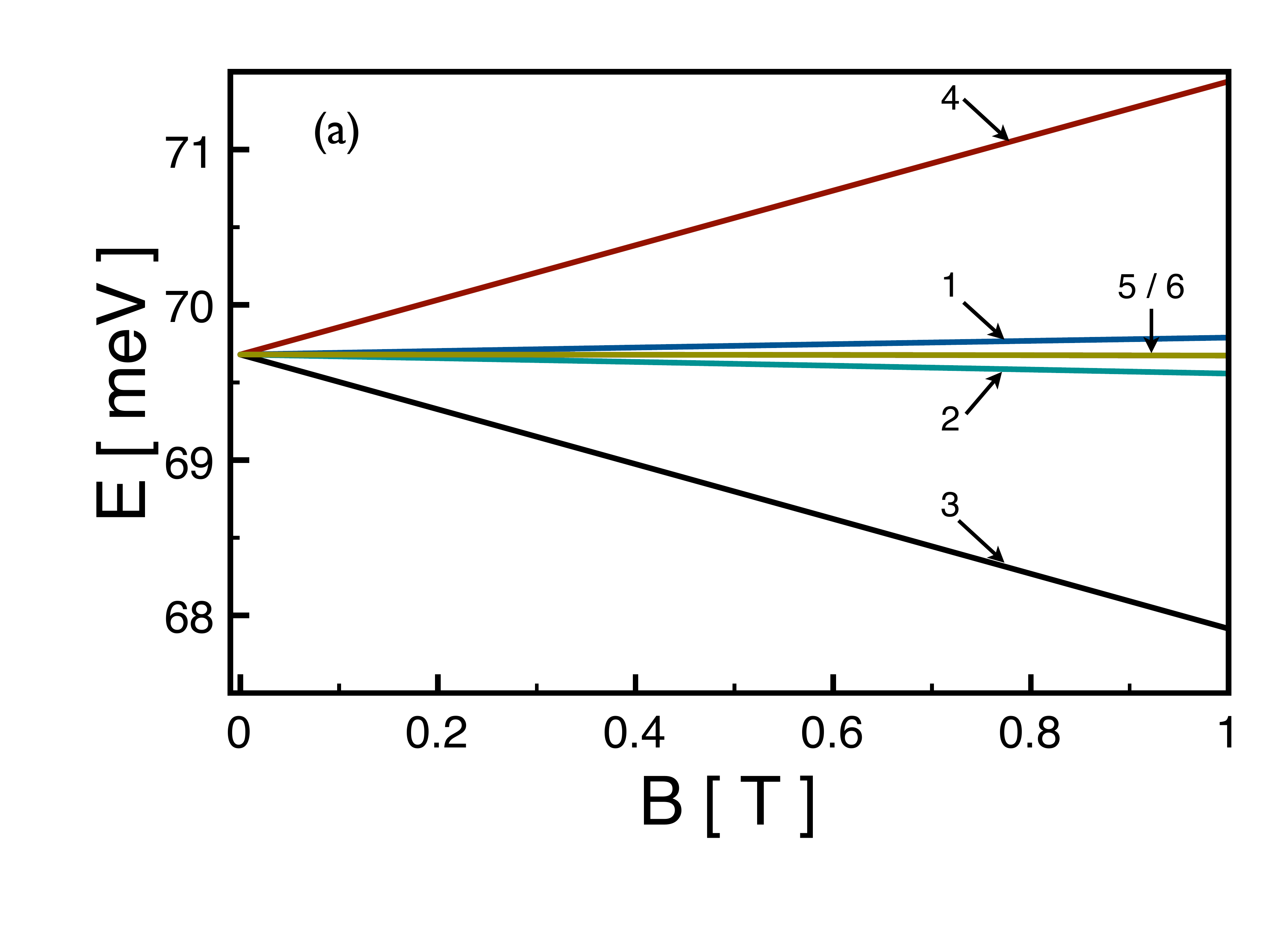}
\end{center}
\end{minipage}
\begin{minipage}[t!]{80mm}
\begin{center}	
\includegraphics[width=80mm]{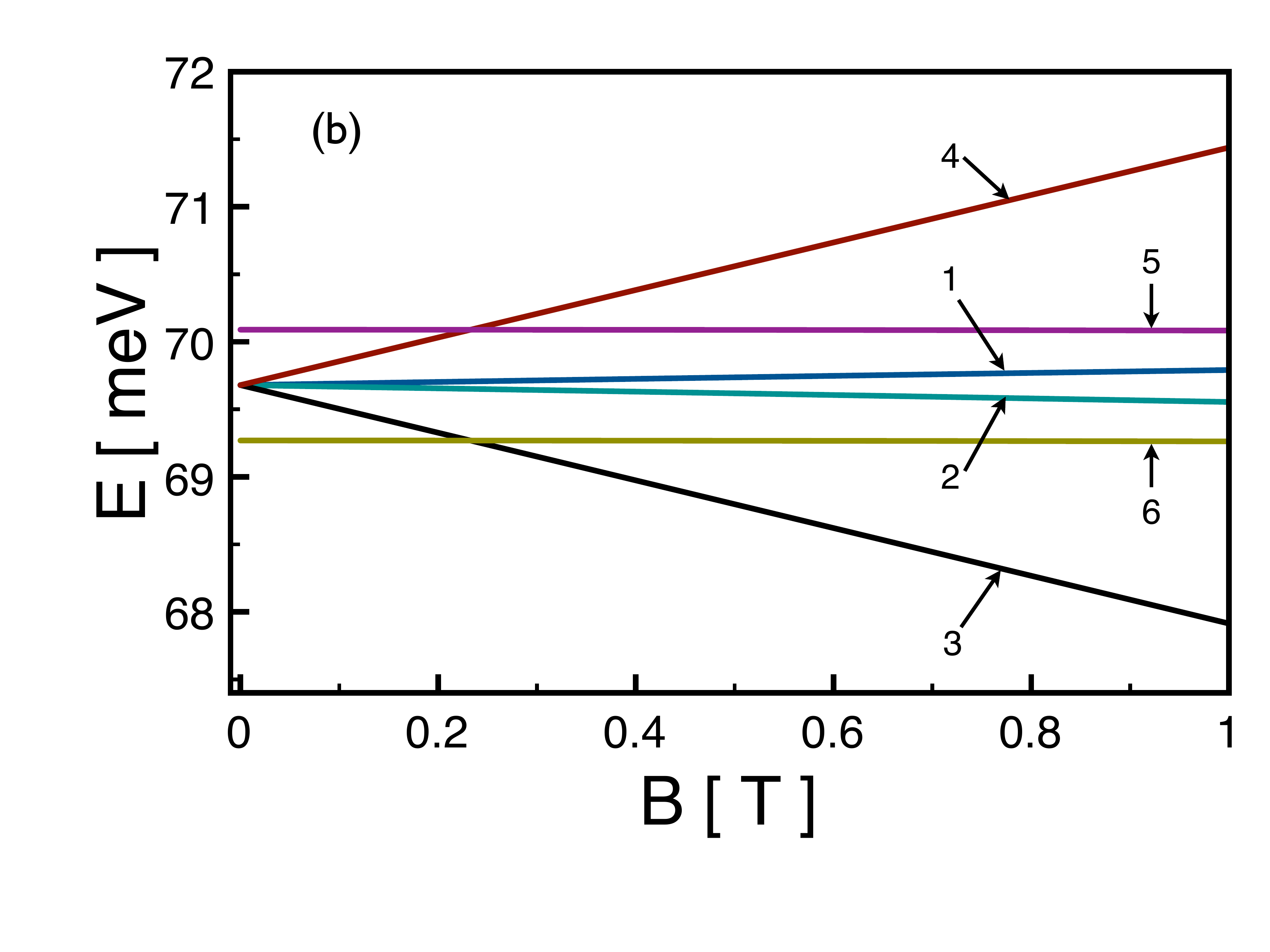}
\end{center}
\end{minipage}
\caption{ \label{02two_el}(color online)
Magnetic field dependence of the energy spectrum of a two-electron single QD. Coulomb interaction is screened by $\kappa=10$; other parameters are the same as in Fig.~\ref{spdqd}. (a) Without SO coupling the spectrum is six-fold degenerate at $B=0$,  and its $B$ dependence originates mostly from the coupling of the orbital and spin magnetic moments to the field. Wavefunctions can be represented as spin singlets--isospin triplets and spin triplets--isospin singlets. (b) With SO coupling the spectrum is split at $B=0$. The level crossing at finite $B$ results in a ground state change from two electrons populating the lower Kramers doublet to two isospin polarized electrons. Numbers with arrows denote the energy that corresponds to a particular state among $\mid\Phi^{02}_{1,\dots,6}\rangle$.}
\end{figure}

The results for the magnetic field dependence of the energy levels of a two-electron QD are shown in Fig.~\ref{02two_el}. We have diagonalized the two particle Hamiltonian for a single QD, cf.~Ref.~\onlinecite{Bulaev}, in the presence of SO coupling as well as a screened Coulomb interaction, $\kappa=10$. QD parameters are chosen as in Sec.~\ref{DQDsol}. The $B$ dependence of the Coulomb interaction terms $U_{02}$ was taken into account consistently. The comparison of panels (a) and (b) demonstrates the effect of SO coupling. In the absence of SO coupling, Fig.~\ref{02two_el}(a), the hexaplet remains unsplit at $B=0$ because $J_{02}$ exchange integrals are disregarded. For $\Delta_{SO}=0$, a classification of these degenerate states in terms of a spin singlet (isospin triplet) and a spin triplet (isospin singlet) is appropriate. The $B$ dependence of the isospin polarized states $|\Phi^{02}_3\rangle, |\Phi^{02}_4\rangle$ and spin polarized states $|\Phi^{02}_1\rangle, |\Phi^{02}_2\rangle$ originates mostly from their orbital and spin magnetic moments, respectively. In the whole region of magnetic fields, the ground state is isospin polarized with both electrons in the $\tau=1$ state having opposite spins.
SO coupling splits the hexaplet at $B=0$, see Fig.~\ref{02two_el}(b), with both electrons populating the lower Kramers doublet in the lowest state $|\Phi^{02}_6\rangle$. The level crossing at $B\approx 0.3\mbox{ T}$ results in a change of the ground state. This transition might also be seen in the two electron spectrum of Ref.~\onlinecite{Kuemmeth}.  At larger fields, the ground state becomes well separated from all higher states.

Although we do not calculate electron attraction due to their coupling to phonons, we note that an estimate shows that it might become comparable to a screened Coulomb repulsion for $\kappa\agt10$. A more detailed investigation of this contribution is needed.

\section{Two particle spectrum as a function of detuning}
\label{detun}

In recent experiments,\cite{Churchill} the dephasing time
$T_2^*$ of a two particle state was obtained by the following
measurement cycle. First the system is prepared in the  $(02)$
configuration whose ground state is non-degenerate for a
CNT-DQD, Fig.~\ref{02two_el}(b). The doubly occupied right dot
might be considered as a double dot in a strongly detuned state
where the detuning energy $\varepsilon$ compensates the strong
Coulomb repulsion; hence, it is energetically favorable for two
electrons to populate the same dot. When decreasing
$\varepsilon$, the Coulomb repulsion and interdot tunneling
allow pushing one of the electrons to the left dot, and the
system is transferred into the $(11)$ configuration. This
produces an additional degree of freedom, manifesting itself in
a quantum well index $L(R)$, and allowing for 16 states in the
$(11)$ configuration, as compared to 6 states in the $(02)$
configuration. The whole space includes $22$ basis states. The
transfer of the system from the six-fold $(02)$ space to
sixteen-fold $(11)$ space is followed by dephasing due to
different mechanisms, including hyperfine interactions and
isospin scattering. When $\varepsilon$  is increased again,
after a certain separation time $\tau_S$,  the system is
prevented from coming back because not all states from $(11)$
are connected by tunneling to the states in $(02)$. This
generalized Pauli blockade arises from the selection rules both
in spin and isospin. The probability of finding both electrons
again in the right dot depends on $\tau_S$, and measuring the
return probability as a function of $\tau_S$ is used to extract
$T_2^*$, as has been done  for CNT-DQDs with $T_2^*=3.2$
ns.\cite{Churchill} Therefore, $T_2^*$ should strongly depend
on the coupling between the energy levels of the $(02)$ and
$(11)$ subsystems that will be investigated below.

\subsection{$(11)$ configuration}

In this section we construct a basis of the two-particle
Hilbert space of the $(11)$ configuration starting from the
$(02)$ configuration basis and using the corresponding
single-particle wavefunctions of Eq.~(\ref{LRwave}). Consider,
e.g., the state of Eq.~(\ref{ST1}) with spin polarized
functions $\mid\Phi_{s}^{02}\rangle$.  There are two
possibilities, either the first or the second electron can
occupy the right dot,
\begin{eqnarray}
\mid \Phi_{Is}^{11}\rangle&=&\frac{1}{\sqrt{2}}\left(\mid RK^+_s s\rangle_1\mid LK^-_s s\rangle_2- 1\rightleftharpoons2\right),\nonumber\\
\mid \Phi_{\tilde{Is}}^{11}\rangle&=&\frac{1}{\sqrt{2}}\left(\mid LK^+_s s\rangle_1\mid RK^-_s s\rangle_2-1\rightleftharpoons2\right).
\label{intermediateI}
\end{eqnarray}
This procedure, when applied to the states  $\mid\Phi_{1,\dots, 6}^{(02)}\rangle$ of Eqs.~(\ref{ST1}), (\ref{calT}) and (\ref{calK}),
results in twelve states of the $(11)$ configuration. From those twelve states we construct combinations that are symmetric and antisymmetric in $L/R$ space.
For example, from Eq.~(\ref{intermediateI}) we find for $s=\uparrow$
\begin{equation}
\mid\Phi_{1\pm}^{11}\rangle\equiv\frac{1}{\sqrt{2}}\left(\mid\Phi_{I\uparrow}^{11}\rangle\pm\mid\Phi_{\tilde{I\uparrow}}^{11}\rangle\right),
\label{from1}
\end{equation}
We denote by $\mid \Phi^{11}_{1+,\dots,6+}\rangle$ and $\mid\Phi^{11}_{1-,\dots,6-}\rangle$, respectively, the symmetric and antisymmetric superpositions in $L/R$ space. In addition, in the $(11)$ configuration there are four states polarized in both spin and isospin
\begin{equation}
\mid\Phi_{i}^{11}\rangle=\frac{1}{\sqrt{2}}\left(\mid LK_s^\tau s \rangle_1\mid RK_s^\tau s\rangle_2 -1\leftrightharpoons2\right).
\label{forbid02}
\end{equation}
All of them are antisymmetric; similar combinations in (02) are forbidden by the Pauli exclusion principle. We use the convention $i=13$ for $\tau=+,s=\uparrow$, $i=14$ for $\tau=+,s=\downarrow$, $i=15$ for $\tau=-,s=\uparrow$, and $i=16$ for $\tau=-,s=\downarrow$.

\subsection{Coulomb and tunneling matrix elements}
\label{sec:CME}

We need to calculate a $16\times16$ matrix of the Coulomb
interaction in the $(11)$ configuration that is similar to
Eq.~(\ref{VCmat}), as well as one- and two-particle matrix
elements that connect the $(02)$- and $(11)$-subspaces.

In the $(11)$ subspace, the matrix of direct Coulomb terms is
diagonal and its matrix elements are equal for symmetric and
antisymmetric combinations. Hence, we need to compute only six
independent matrix elements for the twelve symmetric and
antisymmetric states $|\Phi^{11}_{1\pm,\dots, 6\pm}\rangle$.
They are denoted as $U^+_{11},
U^-_{11},U^\uparrow_{11},U^\downarrow_{11},U^u_{11},U^l_{11}$
according to the spins and isospins of the states involved.
There are four additional Coulomb terms for the states
$\mid\Phi^{11}_{13,14,15,16}\rangle$ that are both spin and
isospin polarized. We denote their Coulomb matrix elements as
$U_{11}^{\uparrow+},U_{11}^{\downarrow+},U_{11}^{\uparrow-},U_{11}^{\downarrow-}$.
Since we have chosen the wavefunctions in such a way that they
are real in the classically forbidden regions (see
Sec.~\ref{DQDsol}) Coulomb matrix elements between the states
$\mid\Phi_{1+,\dots,6+}\rangle$ and
$\mid\Phi_{1-,\dots,6-}\rangle$ vanish. Furthermore, Coulomb
integrals between $|\Phi_{1\pm,\dots,6\pm}\rangle$  and
$|\Phi_{13,14,15,16}\rangle$ vanish because of the spin and
isospin selection rules.

The strongest $B$-dependence occurs for the six matrix elements
$(U_{11}^{+},U_{11}^{-},U_{11}^{\uparrow+},U_{11}^{\downarrow+},U_{11}^{\uparrow-},
U_{11}^{\downarrow-})$ corresponding to the isospin polarized
states (not shown here). Similarly to the $(02)$ configuration,
we attribute this behavior to the $B$-dependence of $k_c$ of
Eq.~(\ref{eqk}). However, comparing to the $(02)$ configuration
(see Fig.~\ref{U02ofB}) the slopes for
$\mid\Phi^{02}_{3}\rangle$ and $\mid\Phi^{11}_{3\pm}\rangle$
have opposite signs. The same holds also for
$\mid\Phi^{02}_{4}\rangle$ and $\mid\Phi^{11}_{4\pm}\rangle$.

In addition to the matrix of the direct Coulomb interaction
discussed above, there are also exchange matrix elements. As
distinct from the (02) configuration, in the $(11)$
configuration there exist a number of interdot exchange matrix
elements that are not annihilated by the requirements of the
spin conservation and the selection rule of Eq.~(\ref{tauCon}).
They include the overlapping densities between the left and
right dot. Specifically, for $\mid\Phi_{1+}^{11}\rangle$,
\begin{equation}
J_{LR}^\uparrow
=\langle LK^-_{\uparrow,2};RK^+_{\uparrow,1}\mid V_C\mid LK^+_{\uparrow,1};RK^-_{\uparrow,2}\rangle.
\label{interdotex}
\end{equation}
The generalization for other states is straightforward. We find
that there are $10$ independent $J$-exchange matrix elements in
the $(11)$ configuration. All of them are positive, and they
are one order of magnitude smaller than the Coulomb terms for
the parameters values chosen. In the (11) configuration, these
exchange terms shift the energies of the antisymmetric states
down and the symmetric states up.

Besides the $(11)$ terms in the Hamiltonian matrix calculated
above, the cross-terms that provide a coupling between the
$(11)$ and $(02)$ configurations are of critical importance.
They originate both from the single-electron tunneling
Hamiltonian $H_T$ and from the two-electron Coulomb Hamiltonian
$V_C$ and connect all $|\Phi^{02}\rangle$ states with the first
twelve $|\Phi^{11}\rangle$ states. The last four
$|\Phi^{11}\rangle$ states of Eq.~(\ref{forbid02}) cannot
tunnel to the $(02)$ configuration by construction.

The first contribution is similar to Eq.~(\ref{TLSham}), yielding twelve matrix elements
\begin{eqnarray}
\langle \Phi_s^{02}\mid H_T\mid \Phi_{s\pm}^{11}\rangle&=&\frac{1}{\sqrt{2}}\left(\eta_{+,s}\pm\eta_{-,s}\right),\nonumber\\
\langle \Phi_\tau^{02}\mid H_T\mid \Phi_{\tau\pm}^{11}\rangle&=&\frac{1}{\sqrt{2}}\left(\eta_{\tau,\uparrow}\pm\eta_{\tau,\downarrow}\right),\nonumber\\
\langle \Phi_u^{02}\mid H_T\mid \Phi_{u\pm}^{11}\rangle&=&\frac{1}{\sqrt{2}}\left(\eta_{-,\downarrow}\pm\eta_{+,\uparrow}\right),\nonumber\\
\langle \Phi_l^{02}\mid H_T\mid \Phi_{l\pm}^{11}\rangle&=&\frac{1}{\sqrt{2}}\left(\eta_{+,\downarrow}\pm\eta_{-,\uparrow}\right).
\label{tunam}
\end{eqnarray}
Here $H_T$ are sums of the tunnel Hamiltonians of
Eq.~(\ref{Htun}) over both electrons, and $\eta_{\tau,s}$ are
defined by Eq.~(\ref{tuneq}). Note that $\pm$ signs in
Eq.~(\ref{tunam}) correspond to symmetric and antisymmetric
states in the $(11)$ configuration. At $B=0$, antisymmetric
combinations $|\Phi^{11}_{5-, 6-}\rangle$  are forbidden from
tunneling to $(02)$ because of spin and isospin conservation
and the relation $\eta_{\tau,s}(B)=\eta_{-\tau,-s}(-B)$
established in Sec.~\ref{DQDsol}. From Eq.~(\ref{tunam})
follows that antisymmetric combinations are not entirely
forbidden from tunneling to the $(02)$ configuration because of
the $(\tau,s)$ dependence of the tunneling matrix elements
$\eta_{\tau,s}$. However, as one concludes from
Fig.~\ref{tunnel}, this dependence is rather weak, only about
1\%, and therefore transitions to $(02)$ states from
antisymmetric $(11)$ states are strongly suppressed. It is in
this sense that we discuss the left-right $(L/R)$ symmetry
selection rules in what follows. Since they are approximate,
the corresponding level crossings transform into narrow avoided
crossings.

The second contribution originates from the Coulomb interaction
and is also represented by 12 matrix elements
\begin{eqnarray}
\langle \Phi_s^{02}\mid V_C\mid \Phi_{s\pm}^{11}\rangle&=&J_{02\leftrightarrow 11}^{s\pm},\nonumber\\
\langle \Phi_\tau^{02}\mid V_C\mid \Phi_{\tau\pm}^{11}\rangle&=&J_{02\leftrightarrow 11}^{\tau\pm},\nonumber\\
\langle \Phi_u^{02}\mid V_C\mid \Phi_{u\pm}^{11}\rangle&=&J_{02\leftrightarrow 11}^{u\pm},\nonumber\\
\langle \Phi_l^{02}\mid V_C\mid \Phi_{l\pm}^{11}\rangle&=&J_{02\leftrightarrow 11}^{l\pm}.
\label{0211coul}
\end{eqnarray}
 All of them include overlap densities from the right and left dots. As an example, we present the exchange integral between the states $|\Phi_{1}^{02}\rangle$ and $|\Phi_{1\pm}^{11}\rangle$
\begin{eqnarray}
J_{02\leftrightarrow 11}^{\uparrow\pm}&=&\frac{1}{\sqrt{2}}\left(\langle RK^-_{\uparrow;2}RK^+_{\uparrow;1}\mid V_C \mid RK^+_{\uparrow;1}LK^-_{\uparrow;2}\rangle\right.\nonumber\\
&\pm&\left.\langle RK^-_{\uparrow;2}RK^+_{\uparrow;1}\mid V_C\mid LK^+_{\uparrow;1}RK^-_{\uparrow;2} \rangle\right).
\label{1102coulex}
\end{eqnarray}
Other exchange integrals of Eq.~(\ref{0211coul}) have a similar structure and the $\pm$ signs
refer to the symmetric and antisymmetric linear combinations
in the $L/R$ degree of freedom for the $(11)$ configuration.
The integrals of Eqs.~(\ref{tunam}) and (\ref{1102coulex}) are subject to the same spin/isospin selection rules and contribute additively to all (avoided) crossings between the $(02)$ and $(11)$ states. Comparing Eqs.~(\ref{interdotex}) and (\ref{1102coulex}) one notices immediately that matrix elements including an odd (even) number of the wave functions of the left or the right dot have opposite (same) signs in the symmetric and antisymmetric states.

\subsection{Energy spectrum}
\label{sec:plots}

For electron spin dynamics as well as for electron manipulation
by gate voltages, the dependence of the energy levels on the
detuning $\varepsilon$ between the two dots and the magnetic
field $B$ is important. Especially, the position of the energy
levels and the width of the avoided crossings that appear due
to tunneling and exchange integrals might be observed in
transport experiments on CNT-DQDs.

\begin{figure}[ht!]
\begin{center}	
\includegraphics[width=80mm]{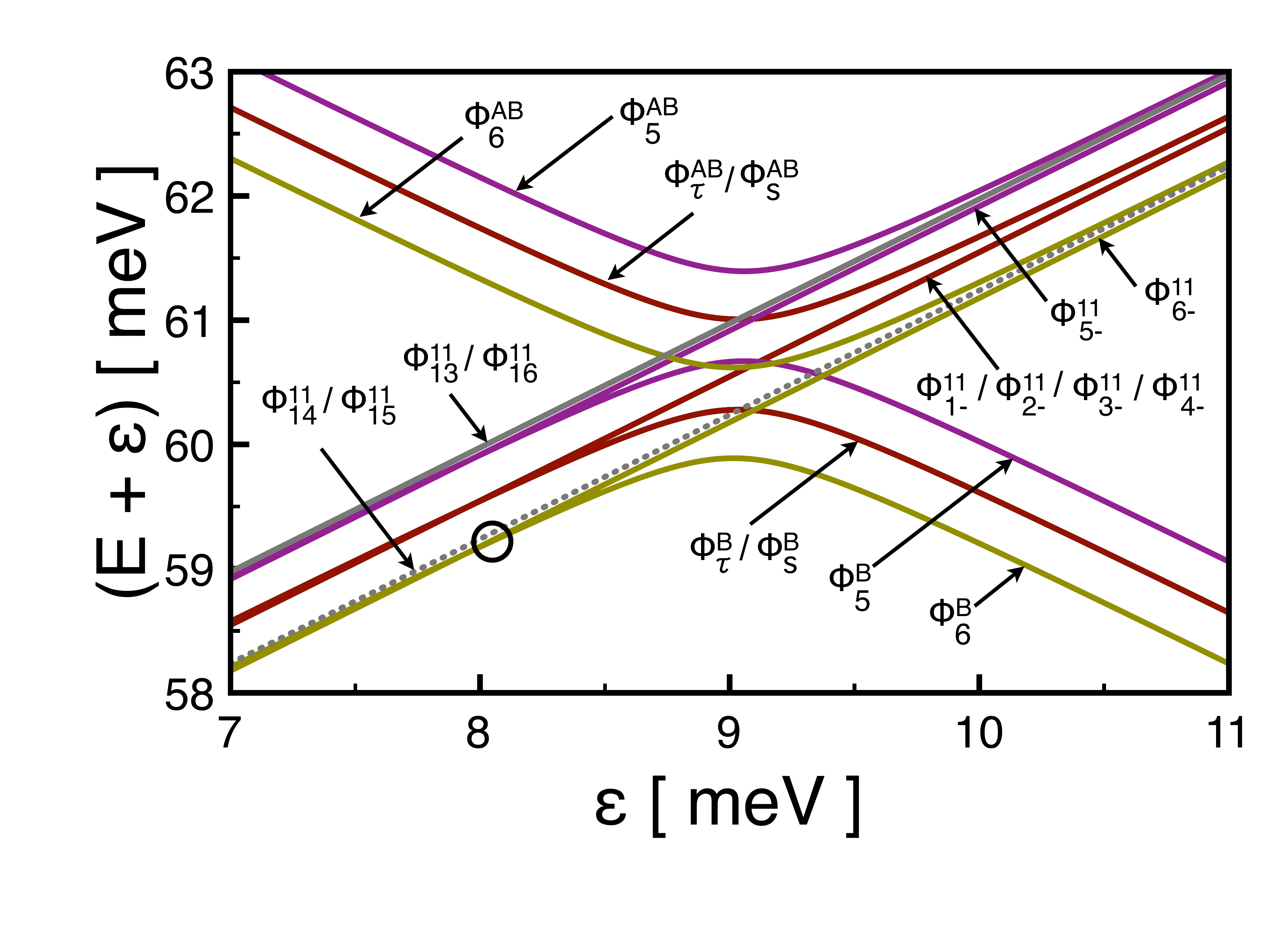}
\caption{ \label{detB0}
(color online) Two particle spectrum at $B=0$ as a function of detuning $\varepsilon$ demonstrating a gradual transition between the $(11)$ and $(02)$ configurations. Parameter values are the same as in Fig.~\ref{spdqd}(b), dielectric constant $\kappa=10$. Hybridized bonding and antibonding states are designated
as $\Phi^{B}_i=\alpha_i\Phi_{i+}^{11}+\beta_i\Phi_i^{02}$ and $\Phi_i^{AB}=\alpha_i\Phi_{i+}^{11}-\beta_i\Phi_i^{02}$, respectively. The coefficients
$\alpha_i, \beta_i>0$ depend on $\varepsilon$ and were found from
numerical diagonalization. Antisymmetric $|\Phi^{11}\rangle$ states that practically do not hybridize with $|\Phi^{02}\rangle$ states are designated as $|\Phi^{11}_{i-}\rangle$. Slashes indicate the states that are either exactly (Kramers) or nearly degenerate; all of them are spin (isospin) polarized. The states $|\Phi_5\rangle$ and $|\Phi_{6}\rangle$ are split by  SO coupling. Remarkably, the ground states of $(02)$ and $(11)$ are not tunnel coupled, and the circle highlights the crossing point between the states $|\Phi^B_6\rangle$ and $|\Phi^{11}_{6-}\rangle$}.
\end{center}
\end{figure}

Fig.~\ref{detB0} presents the result of the two particle
spectrum as a function of detuning at $B=0$. Other dot
parameters are the same as in Fig.~\ref{spdqd}(b). In the lower
right corner, the system is in the $(02)$ configuration and the
ground state is given by $\mid\Phi^{02}_6\rangle$ with both
electrons populating the lower Kramers doublet. The next group
of states ($|\Phi^{02}_s\rangle$ with $s=\uparrow,\downarrow$
and $|\Phi^{02}_\tau\rangle$ with $\tau=\pm$) originates from
the mixed populations of the two Kramers doublets. The
splittings between their energies are controlled by the matrix
elements $U_{02}^\uparrow$ and $U_{02}^+$, the first of which
is small and the second vanishes at $B=0$, see
Fig.~\ref{U02ofB}.  They are not resolved in Fig.~\ref{detB0}.

When $\varepsilon$ decreases, the six $(02)$ states hybridize
with their $|\Phi^{11}_{i+}\rangle$ counterparts  and form
lower and upper (bonding and antibonding) tunnel components,
indicated by B and AB superscripts in the figure. The
$(11)-(02)$ degeneracy point is located at $\varepsilon\approx
U_{02}/2$. Here $U_{02}$ is a mean value of the integrals
$U_{02}^i$ with $i=(\uparrow,\downarrow,+,-,u,l)$ defined in
Sec.~\ref{CoulEl} that differ only within 10\% among each
other. More accurate positions of the degeneracy points for
specific transitions are
$\varepsilon^0_i=[U_{02}^i-(U_{11}^i+J_{LR}^i)]/2$ where
$J_{LR}^i$ are exchange integrals defined by Eq.~(38);
$J_{LR}^i$ are only about a few percents of $U_{02}$. The
widths of the tunnel doublets at $\varepsilon^0_i$ points can
be found
from Eqs.~\eqref{tunam} and \eqref{1102coulex}, for instance, for the up-spin state $|\Phi_{1}\rangle$ it  equals $\sqrt{2}(\eta_{+\uparrow}+\eta_{-\uparrow})+2J^{\uparrow\pm}_{02\leftrightarrow 11}$. As seen in Fig.~\ref{detB0}, the splitting between bonding and antibonding states becomes large in the vicinity of the  $(11)$-$(02)$ degeneracy point and competes with the SO induced splitting.

We note that these equations do not involve the states
$\mid\Phi^{11}_{13,14,15,16}\rangle$ since they are completely
decoupled from all different states and pass through the whole
region of the $(02)-(11)$ resonances without any avoided
crossings.

Remarkable properties of the ground state deserve a special
attention. In the absence of Coulomb interaction (or for very
large $\kappa$), the ground state of the (11) configuration is
controlled by tunneling and is a bonding state that is always
symmetrical, see Fig.~\ref{wavefun}(a). However, because of the
competition between tunneling, Eq.~(\ref{tunam}), and exchange,
Eqs.~(\ref{interdotex}) and (\ref{0211coul}), the ordering of
levels can change. This reordering of levels is a real
consequence of our calculations despite the large value of
$\kappa=10$. The level of the symmetrical hybridized state of
$|\Phi^{11}_{6+}\rangle$ and $|\Phi^{02}_6\rangle$ (designated
as a bonding state $|\Phi^B_6\rangle$ in Fig.~\ref{detB0})
crosses the level of the antisymmetric state
$|\Phi^{11}_{6-}\rangle$ at the point highlighted by a circle
in Fig.~\ref{detB0}. While to the left from the circle these
levels nearly merge in Fig.~\ref{detB0}, they are well resolved
in Fig.~\ref{11two_el}. Under such conditions, the ground
states on the left and right from the $(11)$-$(02)$ resonance
are not connected because of the highly unusual order in which
the levels follow on the left, i.e., in the mostly (11)
configuration. There the antisymmetric state lies below the
symmetric one, as a consequence of the fact that the exchange
integrals (see Sec.~\ref{sec:CME}) prevail over the competing
contribution of the tunneling matrix elements.

Besides the state $|\Phi^{11}_{6-}\rangle$ there are different
(nearly) unconnected states showing up as lines monotonously
increasing with $\varepsilon$ in Fig~\ref{detB0}. Altogether,
there are $10$ states in $(11)$ from which electrons cannot
tunnel to $(02)$, six antisymmetric states and four states
$|\Phi^{11}_{13,14,15,16}\rangle$ which have no counterparts in
the $(02)$ configuration.

In Fig.~\ref{detB0}, there are two kinds of level crossings.
All crossings related to $|\Phi^{11}_{13,14,15,16}\rangle$ are
robust, in the framework of our scheme, against small
perturbations because these are the only states that are both
spin and isospin polarized. In particular, they do not rely on
the $L/R$ symmetry. Distinct from them, the crossings involving
$|\Phi^{11}_{5,6-}\rangle$ states and narrow anticrossings
involving other $|\Phi^{11}_{i-}\rangle$ states rely on the
$L/R$ symmetry (that is not exact and is based on the weak
dependence of $\eta_{\tau,s}$ on $(\tau, s)$ and similar
properties of exchange integrals, cf. Sec.~\ref{detun} B). A
violation of this symmetry transforms them into avoided
crossings, therefore, the widths of the anticrossings can be
controlled by the gates. This might allow control of the system
passage across the point indicated by a circle in
Fig.~\ref{detB0} when performing the $(02)$ to $(11)$
excursions.  Note, the above statements relate only to the
stability of the crossings. The very fact of the existence of
specific crossings depends on the relative magnitude of a
number of different Coulomb and tunnel matrix elements.

One general comment regarding the ground states of two-electron
DQDs is relevant. According to the Lieb-Mattis
theorem,\cite{LM,AM} the ground state of a two-electron system,
at $B=0$ and in the absence of SO coupling, is always a spin
singlet. The proof of this statement (attributed to Wigner in
Ref.~\onlinecite{LM}) is applicable only to scalar wave
functions. Therefore, it is not applicable to carbon nanotubes
where the wave functions are spinors in the pseudospin space.
Also, the classification of the quantum states of SO coupled
systems is generically impossible in terms of the spin
eigenstates. Hence, it is quite remarkable that despite all
these odds, both GS wave functions of Fig.~9 belong to the
group of functions with zero mean value of the spin. A specific
GS function is chosen by a number of competing parameters.

\begin{figure}[ht!]
\begin{center}	
\includegraphics[width=80mm]{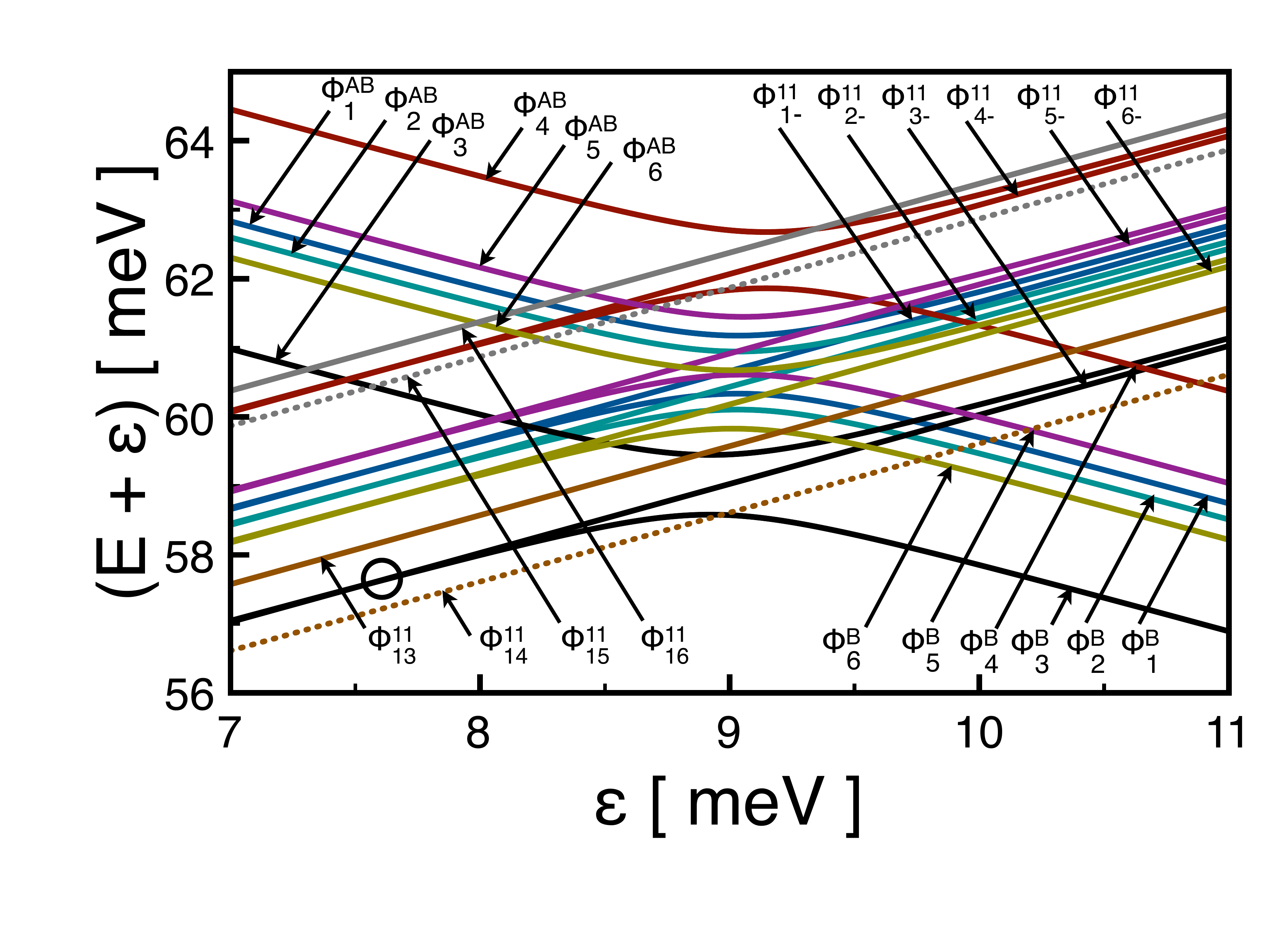}
\caption{ \label{detfinB} (color online) Same as in Fig.~\ref{detB0} for
a magnetic field $B=1$ T. Again, bonding and antibonding states are denoted as $\Phi_i^{B}$ and $\Phi_i^{AB}$, respectively. All degeneracies are removed.
Four $|\Phi^{11}_{13-16}\rangle$ states and six $|\Phi^{11}_{i-}\rangle$ that practically do not hybridize with $|\Phi^{02}\rangle$ states are shown as ascending lines. The circle highlights the intersection of the $|\Phi_3^B\rangle$ hybridized state with $|\Phi^{11}_{3-}\rangle$.  See text for details.}
\end{center}
\end{figure}

\begin{figure}[th!]
\begin{minipage}[t!]{80mm}
\begin{center}	
\includegraphics[width=80mm]{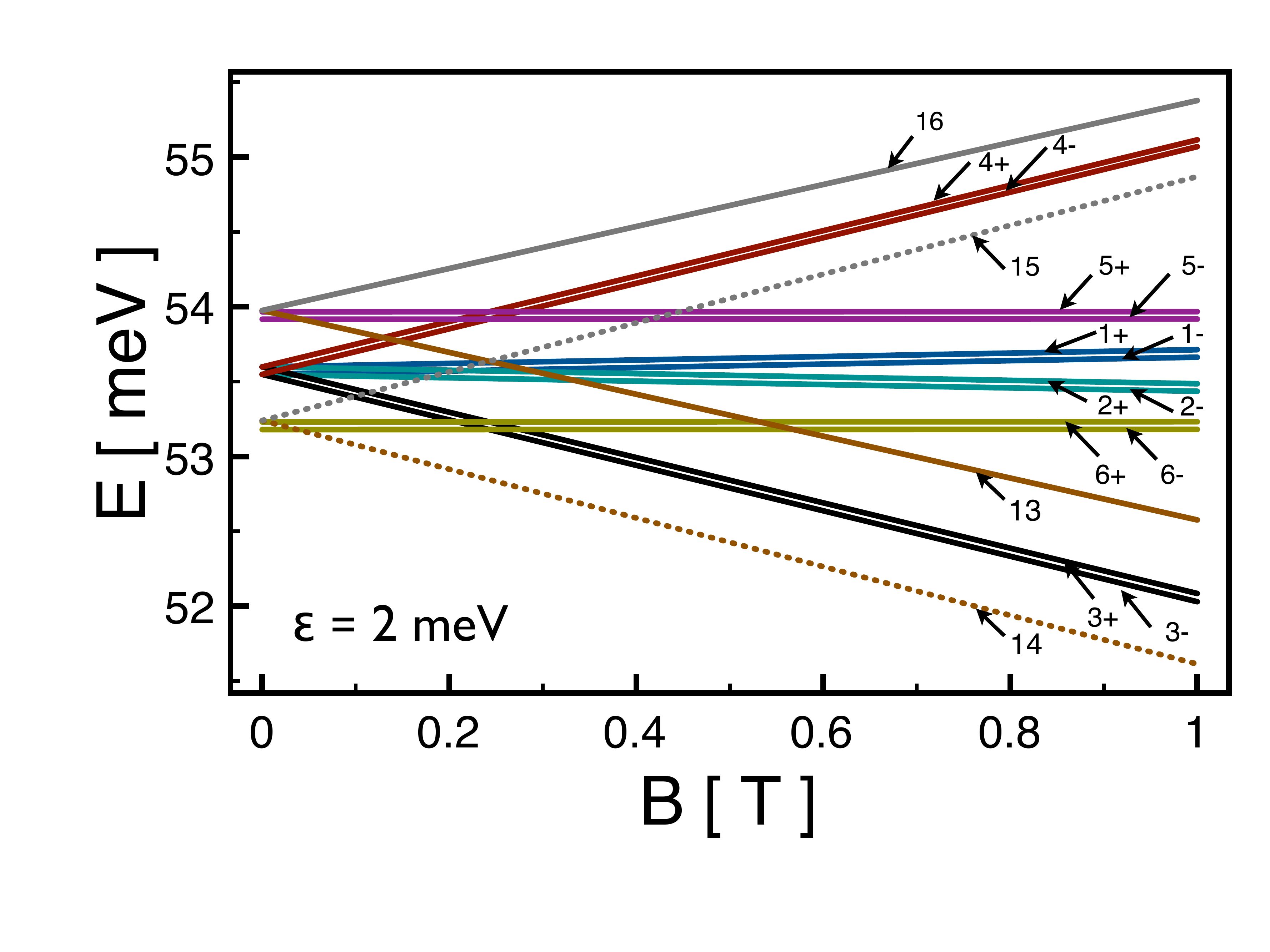}
\end{center}
\end{minipage}
\caption{ \label{11two_el}(color online) Magnetic field dependence of the energy spectrum of a two-electron double dot for $\varepsilon=2$ meV where the admixture of both $(02)$ and $(20)$ configurations is negligibly small. The Coulomb interaction is screened by $\kappa=10$ and other parameters are the same as in Fig.~\ref{spdqd}.
Numbers with arrows denote particular $\mid\Phi^{11}_{1\pm,\dots,6\pm, 14,\dots,16}\rangle$ states. Relative magnitudes of the different level splittings originating from the tunneling and Coulomb interaction are distinctly seen. At $B=0$, the dominating splitting comes from $\Delta_{SO}$ depending on whether one or both electrons belong to the upper or lower Kramers doublet; other contributions are smaller by one order of magnitude. The strong $B$ dependence is controlled by the isospin through $\mu_{orb}$ and a weaker one by the spin through $\mu_B$. The splittings of bonding and antibonding levels are weak, and their sign is controlled by the prevalence of the Coulomb contribution over tunneling.} \end{figure}

Fig.~\ref{detfinB} presents the energy spectrum as a function
of detuning $\varepsilon$ for a magnetic field $B=1$ T. All
degeneracies, both in $(11)$ and $(02)$, are lifted. This field
is large enough to change the symmetry of the ground states
both in the $(11)$- and $(02)$- configurations. Once again, a
GS to GS transition is not allowed. At the $(02)$ side, the
splitting of the isospin doublet $|\Phi_\tau^{02}\rangle$,
$\tau=\pm1$, becomes larger than the SO splitting separating
$|\Phi^{02}_5\rangle$ and $|\Phi^{02}_6\rangle$ states and its
$\tau=+$ component $|\Phi^{02}_3\rangle$ shifts to the spectrum
bottom, in agreement with the experimental findings of
Refs.~\onlinecite{Kuemmeth} and \onlinecite{Churchill}. At the
$(11)$ side, the magnetic field splits the
$|\Phi^{11}_{14,15}\rangle$ Kramers doublet (that was only
slightly above the ground state in Fig.~\ref{detB0}) and shifts
its $\mid\Phi_{14}\rangle$ component to the spectrum bottom; it
is spin and isospin polarized with $s=\downarrow,\tau=+$. In
this context, it is instructive to follow the adiabatic
evolution of the $|\Phi^B_3\rangle$ ground state starting from
$|\Phi^{02}_3\rangle$ in the lower right corner of
Fig.~\ref{detfinB}. After crossing the $|\Phi_{14}\rangle$
ground state (this crossing is both spin conservation and $L/R$
symmetry protected), it passes through a narrow anticrossing
with $|\Phi^{11}_{3-}\rangle$ (protected by the weak $B$
dependence of $\eta_{\tau,s}(B)$ and highlighted by a circle)
to appear only slightly above it as $|\Phi^{11}_{3+}\rangle$,
see Fig.~\ref{11two_el}. Similarly to the related comment to
Fig.~\ref{detB0}, the width of the avoided crossing can be
enhanced by producing asymmetry between the left and right
dots. Since $|\Phi^{11}_{14}\rangle$ and both states
$|\Phi^{11}_{3\pm}\rangle$ possess the same pseudospin $\tau=+$
while $|\Phi^{11}_{3\pm}\rangle$ are spin unpolarized and
$|\Phi^{11}_{14}\rangle$ is spin polarized, the relaxation from
$|\Phi^{11}_{3\pm}\rangle$ to the ground state is only possible
due to the spin nonconservation. Therefore, excursions from
$(02)$ to $(11)$ can be used for measuring the spin relaxation
rate.

We have checked the behavior of the level crossings,
highlighted by circles in Figs.~\ref{detB0} and \ref{detfinB},
when we change the size of the gap $E_g$. With $E_g$ increasing
twice, both crossings remain stable and move to the right,
nearly half way to the $(11)$-$(02)$ degeneracy point
($\varepsilon\sim11$ meV).

In Figs.~\ref{detB0} and \ref{detfinB}, in the vicinity of the
$(11)$-$(02)$ degeneracy point, gross features are dominated by
the $|\Phi^{11}\rangle$ - $|\Phi^{02}\rangle$ hybridization. To
illuminate different properties of the spectrum, its
SO-coupling and $B$-dependence, we present in
Fig.~\ref{11two_el} the energy spectrum at $\varepsilon=2$ meV.
While it was found by the same procedure as Figs.~\ref{detB0}
and \ref{detfinB}, we checked that it is very close to the
spectrum found in the $16\times16$ basis of $|\Phi^{11}\rangle$
functions; in particular, all levels follow in the same order.
This proves that the contribution of the polar configuration
$(20)$ (with both electrons on the left dot) not included in
our calculations is small at $\varepsilon=2$ meV and has only
minor effect on the results.

At $B=0$, the spectrum is dominated by the splitting
originating from the one- or two-fold population of the upper
and lower Kramers doublets separated by $\Delta_{SO}\sim0.4$
meV. Splittings from the inter-dot exchange matrix elements are
lesser: $J_{LR}^{i}\approx 0.03$ meV for $\kappa=10$,
Fig.~\ref{11two_el}.  The tunneling matrix elements $\eta\sim
0.02$ meV [see Fig.~\ref{tunnel} and Eq.~\eqref{tunam}] also
induce lesser splittings. Therefore, the gross structure of the
energy spectrum is controlled by $\Delta_{SO}$, and this
suggests describing it primarily in terms of Kramers doublets
rather than independent spin and isospin populations. The fine
structure inside each group (4+8+4), originating from tunneling
and Coulomb terms, should be accessible for experimental
resolution. One can also distinguish energy differences between
the bonding and antibonding $|\Phi^{11}_{i\pm}\rangle$ states
and the states forbidden for tunneling to $(02)$. It is seen
that the energies of antisymmetric states are lower than the
energies of the corresponding symmetric states in the whole
range of magnetic fields. This is the result of the Coulomb
interaction that favors antisymmetric states prevailing over
tunneling that favors symmetric states.

The $B$-dependence is dominated by the isospin Zeeman coupling
because $\mu_{orb}\gg\mu_B$. However, more careful examination
allows distinguishing differences in the slopes of the states
with the spin and isospin polarized in the same or in opposite
directions, e.g., $|\Phi^{11}_{15,16}\rangle$. A Zeeman
splitting of spin polarized states
$|\Phi^{11}_{1\pm,2\pm}\rangle$ is distinctly seen.

\section{Summary and discussion}
\label{conclusion}

We have studied the detailed structure of the energy spectrum
of a symmetric carbon nanotube double quantum dot with either
one or two electrons confined by an electrostatic potential. We
focused on narrow-gap coated nanotubes allowing efficient gate
control for electronic and spintronic applications, and
investigated the effect of both SO coupling constants,
$\Delta_0$ and $\Delta_1$, on the energy spectrum. The large
effective dielectric constant of such nanotubes
($\kappa\sim10$) in conjunction with a small electron effective
mass $m^*\approx E_g/v^2$ suppresses admixture of the higher
longitudinal modes and allows studying the fine structure of
the spectrum originating from the spin and isospin degrees of
freedom in the framework of a single-mode theory. The
importance of such a study is called for by the experimental
discovery\cite{Churchill2,Churchill} of very narrow features
($\alt10$ mT) in the magnetotransport spectra of DQDs. While a
recent theory\cite{Burkard} proposed a mechanism for developing
magnetocurrent minima with a width
$\Delta_{SO}/\mu_{orb}\sim100$ mT based on the global width
$\Delta_{SO}$ of the SO split spectrum, unveiling the nature of
the narrow features seem to require mechanisms involving
specific quantum levels.  Note that the basic elements of our
theory are also applicable to suspended semiconducting
nanotubes as well, but accounting for higher longitudinal modes
might become necessary.\cite{Wunsch,Rontani,Stecher}

After solving a spinor equation for a double square-well
confining potential in the axial direction, we obtained the
single-particle spectrum in the presence of SO interaction. Due
to the coupling between spin and isospin, the four-fold
degeneracy is lifted at zero magnetic field, $B=0$, which
results in two Kramers doublets corresponding to either aligned
or anti-aligned spin and isospin. We note that while the
diagonal $\Delta_0$ and nondiagonal $\Delta_1$ SO coupling
constants combine in the splitting $\Delta_{SO}$ between the
Kramers doublets, $\Delta_1$ contributes independently to the
interdot tunneling rate. As a result, Kramers doublets acquire
different $B$-dependent tunneling rates, Fig.~\ref{tunnel}; we
estimated this difference using the realistic values of
$\Delta_0$ and $\Delta_1$ found from the experimental data of
Ref.~\onlinecite{Kuemmeth}. They can be observed in experiments
on single-electron transport across double dots.

The basis states for a two-electron DQD in the $(02)$
configuration (both electrons on the right dot) include two
spin polarized as well as two isospin polarized functions, and
two functions belonging to the upper and lower Kramers doublet,
respectively. All of them are spin-isospin coupled, and the
Coulomb interaction energies depend both on the spin and
isospin. In the $(11)$ configuration of a symmetric DQD, these
functions generate (by moving a single electron from the right
to the left dot) 12 basis functions of which 6 are symmetric
(bonding) and 6 antisymmetric (antibonding) in the indices of
the left $(L)$ and right $(R)$ dots. Four more states, all
$L/R$ antisymmetric, have no analogs in the $(02)$ space. Only
bonding modes strongly hybridize with $(02)$ states.

Our main result is the energy spectrum of a two-electron DQD,
calculated for the regime of comparable tunneling and SO
energies, shown in Fig.~\ref{detB0} for $B=0$ and in
Fig.~\ref{detfinB} for $B=1$ T as a function of the detuning
$\varepsilon$ between left and right dots. It is discussed in
Sec.~\ref{sec:plots}. Figures \ref{detB0} and \ref{detfinB}
illustrate how fundamentally the isospin degree of freedom and
its coupling to the spin change the spectrum. This change makes
the analysis of the spectrum much more complicated compared to
the spectrum of GaAs DQDs \cite{Hanson} which consists of the
spin singlet and triplet branches alone.

While both the Pauli blockade and dephasing rate are
challenging goals for experimental studies, investigating the
dephasing rate by initializing the system in the $(02)$
configuration and making excursions into the $(11)$
configuration is more tractable from a theoretical point of
view because of a lesser manifold of quantum states whose width
can be controlled by gate potentials. The effect of a magnetic
field on the mutual position of the lower levels that
influences the relaxation rate between them can be inferred
from Figs.~\ref{detB0} and \ref{detfinB} as discussed in
Sec.~\ref{sec:plots}. As distinct from GaAs where the ground
state is a singlet, in nanotube DQDs this is a double-populated
lower Kramers doublet. We have also found that in our parameter
range the ground state in the $(11)$ configuration is
antisymmetric in $L/R$ indices because the Coulomb repulsion
prevails over tunneling. This unique situation results in the
opposite $L/R$ parity of the ground state on both sides of the
$(11)$-$(02)$ degeneracy point, Fig.~\ref{detB0}. Therefore,
low energy excursions into the $(11)$ configuration can probe
the relaxation rate at small energy transfers and indicate the
position of the $L/R$ symmetry point (deviation from it turns
the level crossing into an anticrossing). When the magnetic
field becomes strong enough, Zeeman splitting shifts a spin
polarized state to the bottom of the spectrum. As a result,
ground states on the left and right differ not only in the
$L/R$ symmetry but also in the two-particle spin wavefunction,
Fig.~\ref{detfinB}. Hence, similar excursions can probe the
spin relaxation rate $\tau^{-1}_s$. Moving up in energy should
allow probing higher states of the $(11)$ configuration.

With such a rich energy spectrum, the very notion of the spin
(Pauli) blockade should be generalized,\cite{Churchill}
including both spin and isospin, and the blockade becomes
rather sensitive to the parameters of the system. Therefore, it
is natural that the  blockade has either been
observed\cite{Churchill} or alternatively not
observed\cite{Kouwenhoven} by different experimental groups.
The outcome should strongly depend on populating the different
$(11)$ levels during the initiation phase, mechanisms of the
relaxation and leakage, and the topology of the dense
intertwined net of the energy levels. A significant challenge
is establishing the optimal conditions for achieving the Pauli
blockade.

The pattern of the energy spectrum, which is rather involved
even in the framework of a simple model (Fig.~\ref{detfinB})
should become even more complicated in realistic systems due to
the $\tau$ non-conservation that is usually controlled by
extrinsic mechanisms and, therefore, might be different in the
left and right dots. It can be taken into account either
phenomenologically by including a term $\Delta_{KK'}\tau_1$
into the Hamiltonian,\cite{Churchill,Bulaev,Rudner,Marcus} or
by modeling a short-range disorder.\cite{Burkard} Likewise, the
electron attraction through their coupling to stretching
phonons that can compete with the Coulomb repulsion at
$\kappa\sim10$, cf. Sec.~\ref{sec:EnSp}, is not studied here
and deserves a detailed investigation in the future.

{\bf Note added:} While completing this manuscript, we became
aware of a paper by v.~Stecher et al.\cite{Stecher} on a
related subject. Both studies are complementary, since
Ref.~\onlinecite{Stecher} focuses mostly on the effect of
electronic correlations in suspended wide-gap nanotubes, while
we concentrate on the fine SO structure of the spectra of
coated narrow-gap nanotubes where such correlations are
suppressed.

\acknowledgments We thank C.~M.~Marcus and A.~A.~Reynoso for
valuable discussions and acknowledge financial support from the
Danish Research Council, INDEX (NSF-NRI), IARPA, the US
Department of Defense, and the Harvard Center for Nanoscale
Systems.

\end{document}